\documentclass[epj]{svjour1}

\usepackage{amsmath}
\usepackage{amssymb}
\usepackage{graphicx}
\usepackage{bm}
\usepackage{psfrag}

\begin{document}

\title{
Casimir-Polder interaction of atoms with magnetodielectric bodies
}

\author{Stefan Yoshi Buhmann\inst{1}\thanks{\emph{Electronic address:}
 \texttt{s.buhmann@tpi.uni-jena.de}}
 \and Ho Trung Dung\inst{2}
 \and Thomas Kampf\inst{3}
 \and Dirk-Gunnar Welsch\inst{1}
}                     

\institute{Theoretisch-Physikalisches Institut,
 Friedrich-Schiller-Universit\"{a}t Jena,
 Max-Wien-Platz 1, 07743 Jena, Germany
 \and Institute of Physics, National Center for Sciences and
 Technology, 1 Mac Dinh Chi Street, District 1, Ho Chi Minh city,
 Vietnam 
 \and Fachbereich Physik, Universit\"{a}t Rostock,
 Universit\"{a}tsplatz 3, 18051 Rostock, Germany
}

\date{Received: \today / Revised version: \today}

\abstract{
A general theory of the Casimir-Polder interaction of single atoms
with dispersing and absorbing magnetodielectric bodies is presented,
which is based on QED in linear, causal media. Both ground-state and
excited atoms are considered. Whereas the Casimir-Polder force acting
on a ground-state atom can conveniently be derived from a perturbative
calculation of the atom-field coupling energy, an atom in an excited
state is subject to transient force components that can only be fully
understood by a dynamical treatment based on the body-assisted vacuum
Lorentz force. The results show that the Casimir-Polder force can be
influenced by the body-induced broadening and shifting of atomic
transitions---an effect that is not accounted for within lowest-order
perturbation theory. The theory is used to study the Casimir-Polder
force of a ground-state atom placed within a magnetodielectric
multilayer system, with special emphasis on thick and thin plates as
well as a planar cavity consisting of two thick plates. It is shown
how the competing attractive and repulsive force components related to
the electric and magnetic properties of the medium, respectively,
can---for sufficiently strong magnetic properties---lead to the
formation of 
potential walls and wells.
\PACS{
 {12.20.-m}{Quantum electrodynamics}   
 \and {34.50.Dy}{Interactions of atoms and molecules with surfaces}
 \and {42.50.Nn}{Quantum optical phenomena in absorbing, dispersive,
  and conducting media}
} 
} 

\maketitle


\section{Introduction}
\label{sec1}

It is one of the most surprising consequences of quantum
electrodynamics (QED) that a neutral unpolarized atom will be subject
to a force when placed in the vicinity of neutral unpolarized 
bodies---even when the body-assisted electromagnetic field is in its 
vacuum state. The existence of this force commonly called Casimir-Polder
(CP) force is experimentally well established. Casimir-Polder forces
have been observed via mechanical means using atomic beam scattering
\cite{Raskin69} and transmission \cite{Anderson88} as well as quantum
reflection \cite{Shimizu01}, and via spectroscopic means
\cite{Sandoghdar92}, inter alia frequency modulated selective
reflection spectroscopy \cite{Oria91}. They are crucial for the
understanding of many phenomena in nature such as the adsorption of
atoms or molecules to surfaces \cite{Bruch83} or even the remarkable
climbing skills of some geckoes and spiders \cite{Autumn02,Kesel04}.
Apart from their important role in atomic-force microscopy
\cite{Binnig86}, major applications of CP forces have been found in
the field of atom optics \cite{Adams94}, where they have been used to
construct atomic mirrors \cite{Shimizu02}, which in connection with
evanescent electromagnetic waves can even operate state-selectively
\cite{Balykin88}. 

If the atom is not too close to the surface of any of the bodies,
a theoretical understanding of the CP force can be obtained with\-in
the framework of macroscopic QED. So, Casimir and Polder derived the
force from the atom-field coupling energy calculated in lowest-order 
perturbation theory \cite{Casimir48}, yielding the potential---in
the following referred to as the van der Waals (vdW) potential---from 
which the force can then be derived. This approach first applied
to the case of a ground-state atom placed in front of a perfectly
conducting plate was later extended to excited atoms \cite{Barton74}
as well as to atoms between two perfectly conducting plates
\cite{Walther97}. Moreover, the concept has been used to calculate the
force acting on an atom placed in front of a semi-infinite dielectric
half space \cite{Tikochinski93}, near a carbon nanotube
\cite{Bondarev04}, or between two dielectric plates of finite
thickness \cite{Zhou95}. Recently, the ideas of Casimir and Polder
have been generalized to allow for dispersing and absorbing bodies
\cite{Buhmann04a,Buhmann04b}. In parallel with the exact QED approach,
a semiphenomenological method has been established and widely used.
According to this approach, the coupling energy is expressed in terms
of correlation functions for the atom and/or the electromagnetic field
which in turn are related to susceptibilities via the
dissipation-fluctuation theorem. The result---which in principle
applies to arbitrary geometries---was first used for a ground-state
atom placed in front of a perfectly conducting half space
\cite{McLachlan63}, a dielectric half space \cite{McLachlan63b}, and a
dielectric two-layer system \cite{Wylie84}. Later, atoms in excited
energy eigenstates were included in the concept \cite{Wylie85}.
Effects of surface roughness \cite{Henkel98} and finite temperature
\cite{McLachlan63c,Henkel02} and---in the case of the semi-infinite
half space---different materials such as birefringent dielectric 
\cite{Gorza01} or even magnetodielectric matter \cite{Kryszewski92}
have been studied. 

In the large body of work on forces between polarizable objects the
electric properties of the involved objects have typically been the
focus of interest. Nevertheless, the fact that Maxwell's equations in
the absence of (free) charges and currents are invariant under a
duality transformation between electric and magnetic fields can be
exploited to extend the notion of these forces to magnetically
polarizable objects. Thus, knowing the attractive force between two
electrically polarizable atoms, one can infer the existence of an
analogeous attractive force between two magnetically polarizable
atoms, which may be obtained from the former by replacing the
electric polarizabilities by the corresponding magnetic ones. In
contrast, the force between two atoms of opposed polarizability (i.e.,
electric/magnetic) is repulsive \cite{Sucher70}. While the repulsive
force in the retarded limit obeys the same power law
\cite{Sucher68,Boyer69}, the leading contribution to the repulsive
force in the nonretarded limit is weaker than the corresponding
attractive force between two electrically polarizable atoms by two
powers in the atom-atom-separation \cite{Farina02}. A similar
hierarchy of attractive and repulsive forces with corresponding
asymptotic power laws has been found for the Casimir force between two
semi-infinite half spaces possessing electric or magnetic properties,
respectively \cite{Henkel04} (see also Sec.~\ref{sec4.2}). 

Surprisingly, the CP force between a single atom and a macroscopic
body has not yet been considered in detail in this context. The
repulsive retarded force found for a magnetically polarizable atom
interacting with a perfectly conducting plate implies---by virtue of
a duality transformation---that the retarded force between an
electrically polarizable atom and an infinitely permeable plate
should also be repulsive, which provides interesting opportunities. 
A thorough analysis of the CP force between a single atom and a system
of genuinely magnetodielectric bodies is desirable for three reasons:
First, the availability of sensitive spectroscopic measurement
techniques \cite{Oria91} suggests that in this case an experimental
verification of repulsive forces is much more likely than in the case
of two macroscopic bodies, where the mechanic measurements are
currently restricted to distance regimes of purely attractive forces
\cite{Kenneth02}. Second, the rapidly increasing amount of
miniaturization in current technologies shows that CP-type forces will
have to be thoroughly taken into account in the near future. Even
today, Casimir forces are responsible for the problem of the sticking
of nanodevices common in nanotechnology \cite{Henkel04}, while CP
forces can pose severe limits on the trap lifetime on atom chips
\cite{Lin04}. Third, the recent fabrication of metamaterials with
controllable magnetic and electric properties in the microwave regime
\cite{Pendry99,Smith00} and the rapid developments in this field
imply the question of to what extent CP forces could be shaped by a
clever use of magnetodielectrics with appropriate properties. A
thorough analysis of the dependence of CP forces---including both
perturbing and desirable effects---on relevant material and
geometrical parameters can therefore add further impetus
to the research and design of new materials as well as show the
direction towards which intensified efforts should be aimed---having
in mind the future perspective of CP-force engineering.

In this paper we study---within the frame of exact quantization of the
macroscopic electromagnetic field in linear, causal media (reviewed in
Sec.~\ref{sec2})---the CP interaction between a single atom and an
arbitrary arrangement of linear, dispersing, and absorbing
magnetodielectric bodies. We approach the problem from two sides, by
first considering the perturbative atom-field coupling energy
(Sec.~\ref{sec3.1}), and by second going beyond perturbation theory,
presenting a dynamical approach based on the Lorentz force 
averaged with respect to the body-assisted electromagnetic vacuum and
the internal atomic motion (Sec.~\ref{sec3.2}). Section \ref{sec4} is
then devoted to the particular problem of the competing effects of the
electric and magnetic material properties on the CP force acting on a
ground-state atom placed within a genuinely magnetodielectric
multilayer system, where we study the examples of asymptotically thick
and thin plates (Secs.~\ref{sec4.1} and \ref{sec4.2}, respectively) as
well as a simple planar cavity (Sec.~\ref{sec4.3}) in more detail.
Finally, a summary and some concluding remarks are 
given in Sec.~\ref{sec5}.


\section{QED in dispersing and absorbing magnetodielectric media}
\label{sec2}

The study of the interaction of atoms with the electromagnetic field
in the presence of linearly responding magnetodielectric bodies
requires quantization of the electromagnetic field in linear, causal
media. Consider an arbitrary arrangement of neutral, linear,
isotropic, dispersing, and absorbing magnetodielectric bodies, which
can be characterized by their (relative) electric permittivity
$\varepsilon(\vec{r},\omega)$ and their (relative) magnetic
permeability $\mu(\vec{r},\omega)$. Both quantities are
comp\-lex-valued functions that vary with space and---in accordance
with the Kramers-Kronig relations---with frequency. Note that for
absorbing media we have
$\mathrm{Im}\,\varepsilon(\vec{r},\omega)$ $\!>$ $\!0$ and
$\mathrm{Im}\,\mu(\vec{r},\omega)$ $\!>$ $\!0$. In the absence of free
charges and currents Maxwell's equations in frequency space are given
by
\begin{eqnarray}
\label{eq1}
 \bm{\nabla}\cdot\underline{\hat{\vec{B}}}(\vec{r},\omega) = 0, &\quad&
 \bm{\nabla}\times \underline{\hat{\vec{E}}}(\vec{r},\omega)
 -i\omega\underline{\hat{\vec{B}}}(\vec{r},\omega) = 0, 
\\
\label{eq2}
 \bm{\nabla}\cdot\underline{\hat{\vec{D}}}(\vec{r},\omega) = 0, &\quad& 
\bm{\nabla}\times\underline{\hat{\vec{H}}}(\vec{r},\omega)
 +i\omega\underline{\hat{\vec{D}}}(\vec{r},\omega) = 0,
\end{eqnarray}
where
\begin{eqnarray}
\label{eq3}
\hat{\underline{\vec{D}}}(\vec{r},\omega)
 &=& \varepsilon_0\hat{\underline{\vec{E}}}(\vec{r},\omega)
 +\hat{\underline{\vec{P}}}(\vec{r},\omega),
\\
\label{eq4}
 \hat{\underline{{\bf H}}}({\bf r},\omega)
 &=& \kappa_0\hat{\underline{\vec{B}}}(\vec{r},\omega) 
 -\hat{\underline{\vec{M}}}(\vec{r},\omega)
\end{eqnarray}
($\kappa_0$ $\!=$ $\!\mu_0^{-1}$), and the constitutive relations read
\begin{eqnarray}
\label{eq5}
\hat{\underline{\vec{P}}}(\vec{r},\omega)
 &=&\varepsilon_0[\varepsilon(\vec{r},\omega)-1]
 \hat{\underline{\vec{E}}}(\vec{r},\omega)
 +\hat{\underline{\vec{P}}}_\mathrm{N}(\vec{r},\omega),\\
\label{eq6}
 \hat{\underline{\vec{M}}}(\vec{r},\omega)
 &=&\kappa_0[1-\kappa(\vec{r},\omega)]
 \hat{\underline{\vec{B}}}(\vec{r},\omega) 
 +\hat{\underline{\vec{M}}}_\mathrm{N}(\vec{r},\omega)
\end{eqnarray}
[$\kappa({\bf r},\omega)$ $\!=$ $\!\mu^{-1}({\bf r},\omega)$]. In
Eqs.~(\ref{eq5}) and (\ref{eq6}),
$\hat{\underline{\vec{P}}}_\mathrm{N}(\vec{r},\omega)$ and 
$\hat{\underline{\vec{M}}}_\mathrm{N}(\vec{r},\omega)$ denote noise
polarization and noise magnetization, which are unavoidably associated with
electric and magnetic losses, respectively.
Eqs.~(\ref{eq1})--(\ref{eq6}) imply that the electric field obeys a
Helmholtz equation
\begin{equation}
\label{eq7}
\left[\bm{\nabla}\times\kappa(\vec{r},\omega)\bm{\nabla}\times
 \,-\,\frac{\omega^2}{c^2}\varepsilon(\vec{r},\omega)\right]
 \underline{\hat{\vec{E}}}(\vec{r},\omega)
 =i\omega\mu_0
 \underline{\hat{\vec{j}}}_\mathrm{N}(\vec{r},\omega),
\end{equation}
the source term of which is given by the noise current density
\begin{equation}
\label{eq8}
\hat{\underline{\vec{j}}}_\mathrm{N}(\vec{r},\omega)
 = -i\omega\hat{\underline{\vec{P}}}_\mathrm{N}(\vec{r},\omega)
 +\bm{\nabla} \times
 \hat{\underline{\vec{M}}}_\mathrm{N}(\vec{r},\omega). 
\end{equation}
Upon introducing the (classical) Green tensor, which is defined by
the equation
\begin{equation}
\label{eq9}
\left[\bm{\nabla}\times \kappa(\vec{r},\omega)\bm{\nabla}\times
 \,-\,\frac{\omega^2}{c^2}\,\varepsilon(\vec{r},\omega)\right]
 \tens{G}(\vec{r},\vec{r}',\omega)=\bm{\delta}(\vec{r}-\vec{r}')
\end{equation}
together with the boundary condition
\begin{equation}
\label{eq10}
\tens{G}(\vec{r},\vec{r}',\omega)\to 0 
\quad\mbox{for }|\vec{r}-\vec{r}'|\to \infty,
\end{equation}
the solution to Eq.~(\ref{eq7}) can be given in the form 
\begin{equation}
\label{eq11}
\hat{\underline{\vec{E}}}(\vec{r},\omega)
 =i\omega\mu_0 \int \mathrm{d}^3 r'
 \,\tens{G}(\vec{r},\vec{r}',\omega)
 \cdot\hat{\underline{\vec{j}}}_\mathrm{N}(\vec{r}',\omega).
\end{equation}
The Green tensor has the following useful properties \cite{Ho03}:
\begin{equation}
\label{eq12}
 \tens{G}^{\ast}(\vec{r},\vec{r}',\omega)
     =\tens{G}(\vec{r},\vec{r}',-\omega^{\ast}),
\end{equation}
\begin{equation}
\label{eq13}
     \tens{G}(\vec{r},\vec{r}',\omega)
     =\tens{G}^\top(\vec{r}',\vec{r},\omega),
\end{equation}
\begin{align}
\label{eq14}
&\int\!\mathrm{d}^3 s\, \Bigl\{
 \mathrm{Im}\,\kappa(\vec{s},\omega)\bigl[
 \tens{G}(\vec{r},\vec{s},\omega)\!\times\!
 \overleftarrow{\bm{\nabla}}_{\!\!\vec{s}}\bigr]
 \cdot
 \bigl[{\bm{\nabla}}_{\vec{s}}\!\times\!
 \tens{G}^\ast(\vec{s},\vec{r}',\omega)\bigr]
 \nonumber\\&
 +\frac{\omega^2}{c^2}\, 
 \mathrm{Im}\,\varepsilon(\vec{s},\omega)
 \,\tens{G}(\vec{r},\vec{s},\omega)
 \cdot\tens{G}^\ast(\vec{s},\vec{r}',\omega)\Bigr\}
 =\mathrm{Im}\,\tens{G}(\vec{r},\vec{r}',\omega),
\end{align}
where $\bigl[\tens{G}(\vec{r},\vec{s},\omega)\!\times\!
\overleftarrow{\bm{\nabla}}_{\!\!\vec{s}}\bigr]_{ij}$
$\!=$ $\!\epsilon_{jkl}\partial^s_l G_{ik}(\vec{r},\vec{s},\omega)$. 

Having expressed the electric-field operator in the frequency domain
in the form of Eq.~(\ref{eq11}), quantization can be performed
by relating noise polarization and noise magnetization to Bosonic
vector fields $\hat{\vec{f}}_e(\vec{r},\omega)$ and
$\hat{\vec{f}}_m(\vec{r},\omega)$, 
\begin{eqnarray}
\label{eq15}
\bigl[\hat{f}_{\lambda i}(\vec{r},\omega),
 \hat{f}_{\lambda' j}^\dagger(\vec{r}',\omega')\bigr]
 &=& \delta_{\lambda\lambda'}\delta_{ij}
 \delta(\vec{r}-\vec{r}')\delta(\omega-\omega'),\quad\\
\label{eq16}
\bigl[\hat{f}_{\lambda i}(\vec{r},\omega),
 \hat{f}_{\lambda' j}(\vec{r}',\omega')\bigr]&=&0
\end{eqnarray}
($\lambda$, $\!\lambda'$ $\!\in$ $\!\{e,m\}$), as follows:
\begin{align}
\label{eq17}
&\hat{\underline{\vec{P}}}_\mathrm{N}(\vec{r},\omega)
 =i\sqrt{\frac{\hbar\varepsilon_0}{\pi}
 \mathrm{Im}\,\varepsilon(\vec{r},\omega)}
 \,\hat{\vec{f}}_{e}({\bf r},\omega),\\
\label{eq18}
&\hat{\underline{\vec{M}}}_\mathrm{N}(\vec{r},\omega)
 =\sqrt{-\frac{\hbar\kappa_0}{\pi}
 \mathrm{Im}\,\kappa(\vec{r},\omega)}
 \,\hat{\vec{f}}_{m}(\vec{r},\omega).
\end{align}
Combining Eqs.~(\ref{eq8}), (\ref{eq11}), (\ref{eq17}), and
(\ref{eq18}), on using the convention
\begin{equation}
\label{eq19}
 \hat{O}(\vec{r})=\int_0^{\infty}\mathrm{d}\omega\,
 \hat{\underline{O}}(\vec{r},\omega)+\mathrm{H.c.}\,,
\end{equation}
yields the body-assisted electric field in terms of the 
dynamical variables $\hat{\vec{f}}_\lambda(\vec{r},\omega)$ and
$\hat{\vec{f}}_\lambda^\dagger(\vec{r},\omega)$,
\begin{equation}
\label{eq20}
 \hat{\vec{E}}({\bf r})
 \!=\!\sum_{\lambda=e,m}\int_0^{\infty}\!\mathrm{d}\omega\,
 \int\mathrm{d}^3r'\,
 \tens{G}_\lambda(\vec{r},\vec{r}',\omega)
 \cdot\hat{\vec{f}}_\lambda(\vec{r}',\omega)
 +\mathrm{H.c.}\,,
\end{equation}
where 
\begin{equation}
\label{eq21}
\tens{G}_e(\vec{r},\vec{r}',\omega) 
 =i\,\frac{\omega^2}{c^2}
 \sqrt{\frac{\hbar}{\pi\varepsilon_0}\,
 \mathrm{Im}\,\varepsilon(\vec{r}',\omega)}\,
 \tens{G}(\vec{r},\vec{r}',\omega),
\end{equation}
\begin{equation}
\label{eq22}
\tens{G}_m(\vec{r},\vec{r}',\omega)
 =-i\,\frac{\omega}{c}
 \sqrt{-\frac{\hbar}{\pi\varepsilon_0}\,
 \mathrm{Im}\,\kappa(\vec{r}',\omega)}
 \bigl[\tens{G}(\vec{r},\vec{r}',\omega)\!\times\!\!
 \overleftarrow{\bm{\nabla}}_{\!\!\vec{r}'}\bigr].
\end{equation}
The body-assisted induction field can be obtained by combining
Eqs.~(\ref{eq1}), (\ref{eq8}), (\ref{eq11}), (\ref{eq17}), and
(\ref{eq18}), resulting in 
\begin{equation}
\label{eq23}
\hat{\vec{B}}({\bf r})\!=\!\!\!
 \sum_{\lambda=e,m}\!\int_0^{\infty}\!
 \frac{\mathrm{d}\omega}{i\omega}
 \!\!\int\!\mathrm{d}^3r'\,\bm{\nabla}\times
 \tens{G}_\lambda(\vec{r},\vec{r}',\omega)
 \cdot
 \hat{\vec{f}}_\lambda(\vec{r}',\omega)
 +\mathrm{H.c.}.
\end{equation}
It can be proved \cite{Ho03} that the fundamental (equal-time)
commutation relations 
\begin{align}
\label{eq24}
&\bigl[\hat{E}_i(\vec{r}),\hat{E}_j(\vec{r}')\bigr]=0 
 =\bigl[\hat{B}_i(\vec{r}),\hat{B}_j(\vec{r}')\bigr],\\
\label{eq25}
&\bigl[\varepsilon_0\hat{E}_i(\vec{r}),
 \hat{B}_j(\vec{r}')\bigr]
 =-i\hbar\epsilon_{ijk} \partial_k\delta(\vec{r}-\vec{r}')
\end{align}
are valid. It is obvious that the Hamiltonian of the system consisting
of the electromagnetic field and the bodies can be given by 
\begin{equation}
\label{eq26}
\hat{H}_\mathrm{F}
= \sum_{\lambda=e,m}
\int\mathrm{d}^3 r \int_0^\infty \mathrm{d}\omega\,\hbar\omega\,
\hat{\vec{f}}_{\lambda}^\dagger(\vec{r},\omega)
\cdot\hat{\vec{f}}_{\lambda}(\vec{r},\omega).
\end{equation}

After having thus established a consistent description of the
quantized body-assisted electromagnetic field, one can proceed by
introducing atom-field interactions. To that end, consider a single
neutral atomic system such as an atom or a molecule (briefly referred
to as atom in the following) consisting of particles $\alpha$ with
charges $q_\alpha$ ($\sum_\alpha q_\alpha$ $\!=$ $\!0$), masses
$m_\alpha$, positions $\hat{\vec{r}}_\alpha$, and canonically
conjugated momenta $\hat{\vec{p}}_\alpha$, the dynamics of which can
be described, within the multipolar coupling scheme (cf., e.g.,
Ref.~\cite{Craig84}), by the atomic Hamiltonian
\begin{equation}
\label{eq27}
\hat{H}_\mathrm{A}=\sum_{\alpha}
 \frac{\hat{\vec{p}}_\alpha{\!^2}}{2m_\alpha}
 +\frac{1}{2\varepsilon_0}\int\mathrm{d}^3r\,
 \hat{\vec{P}}_\mathrm{A}^2(\vec{r}).
\end{equation}
Here, 
\begin{equation}
\label{eq28}
\hat{\vec{P}}_\mathrm{A}(\vec{r})=\sum_\alpha q_\alpha
 \hat{\bar{\vec{r}}}_\alpha
 \int_0^1 \mathrm{d}\lambda
 \,\delta(\vec{r}-\hat{\vec{r}}_\mathrm{A}
 -\lambda\hat{\bar{\vec{r}}}_\alpha)
\end{equation}
is the atomic polarization relative to the center of mass
\begin{equation}
\label{eq29}
\hat{\vec{r}}_\mathrm{A}=\sum_\alpha\frac{m_\alpha}{m_\mathrm{A}}
 \,\hat{\vec{r}}_\alpha
\end{equation}
($m_\mathrm{A}$ $\!=$ $\!\sum_\alpha m_\alpha$), where
\begin{equation}
\label{eq30}
\hat{\bar{\vec{r}}}_\alpha=\hat{\vec{r}}_\alpha
 -\hat{\vec{r}}_\mathrm{A}
\end{equation}
denotes relative particle coordinates. In electric dipole
approximation the atom-field interaction can be described by the
Hamiltonian
\begin{equation}
\label{eq31}
\hat{H}_\mathrm{AF}=-\hat{\vec{d}}\cdot
 \hat{\vec{E}}(\hat{\vec{r}}_\mathrm{A})
 +\frac{1}{2m_\mathrm{A}}
 \bigl[\hat{\vec{p}}_\mathrm{A},\hat{\vec{d}}\!\times\!
 \hat{\vec{B}}(\hat{\vec{r}}_\mathrm{A})\bigr]_+
\end{equation}
($[\hat{\vec{a}},\hat{\vec{b}}]_+=\hat{\vec{a}}\cdot\hat{\vec{b}}
+\hat{\vec{b}}\cdot\hat{\vec{a}}$), where
\begin{equation}
\label{eq32}
\hat{\vec{d}}
 =\sum_\alpha q_\alpha\hat{\vec{r}}_\alpha
 =\sum_\alpha q_\alpha\hat{\bar{\vec{r}}}_\alpha
\end{equation}
denotes the electric dipole moment of the atom, 
\begin{equation}
\label{eq33}
 \hat{\vec{p}}_\mathrm{A}=\sum_\alpha \hat{\vec{p}}_\alpha
\end{equation}
is its total (canonical) momentum, and
$\hat{\vec{E}}(\hat{\vec{r}}_\mathrm{A})$ and
$\hat{\vec{B}}(\hat{\vec{r}}_\mathrm{A})$, respectively, are 
given by Eqs.~(\ref{eq20}) and (\ref{eq23}) (for details, see \cite{Ho03}). 
The second term in Eq.~(\ref{eq31}) is known as the R\"{o}ntgen
interaction, it is obviously due to the translational motion of the
atom. Combining Eqs.~(\ref{eq26}), (\ref{eq27}), and (\ref{eq31}), the
total system can be described by the Hamiltonian
\begin{equation}
\label{eq33.1}
\hat{H}=\hat{H}_\mathrm{F}+\hat{H}_\mathrm{A}
+\hat{H}_\mathrm{AF}.
\end{equation}


\section{Casimir-Polder force}
\label{sec3}

The existence of the CP force acting on a neutral, nonpolar atom
placed in the vicinity of neutral, nonpolar bodies---even when the 
body-assisted electromagnetic field is in its vacuum state
$|\{0\}\rangle$ [defined by
$\vec{f}_{\lambda}(\vec{r},\omega)|\{0\}\rangle$ $\!=$ $\!0$]---can
be understood by noting that the vacuum electromagnetic field,
while vanishing on average, exhibits nonzero fluctuations around this
average, which can become high\-ly inhomogeneous due to the 
presence of the bodies. In particular, for the electric field we have
\begin{equation}
\label{eq34}
\langle\hat{\vec{E}}(\vec{r})\rangle=0
\end{equation} 
[cf. Eqs.~(\ref{eq20})--(\ref{eq22})] and 
\begin{align}
\label{eq35}
\langle[\Delta\hat{\vec{E}}(\vec{r})]^2\rangle
&=\langle\{0\}|\hat{\vec{E}}^2(\vec{r})|\{0\}\rangle-
\langle\{0\}|\hat{\vec{E}}(\vec{r})|\{0\}\rangle^2
\nonumber\\
&=\frac{\hbar}{\pi\varepsilon_0}
\int_0^\infty\mathrm{d}\omega\,
\frac{\omega^2}{c^2}\,\mathrm{Im}\bigl[\mathrm{Tr}\, 
\tens{G}(\vec{r},\vec{r},\omega)\bigr]
\end{align}
[combine Eqs.~(\ref{eq20})--(\ref{eq22}) with commutation relations
(\ref{eq15}) and (\ref{eq16}), and use integral equation
(\ref{eq14})]. The inhomogeneous part of the vacuum fluctuations of the
body-assisted electromagnetic field, in combination with the quantum
fluctuations of the atomic electric dipole moment, can be regarded
responsible for the CP force.


\subsection{Perturbative treatment}
\label{sec3.1}

In the perturbative treatment, the CP force acting on an atom in an
energy eigenstate $|l\rangle$ (with corresponding energy $E_l$) 
is commonly derived from the lowest-order energy shift
$\Delta E_l$ of the state $|l\rangle|\{0\}\rangle$ due to the
(electric part of the) interaction Hamiltonian (\ref{eq31}), which is
a good approximation provided that both the internal and the
center-of-mass motion of the atom are nonrelativistic. The
position-dependent part of the energy shift is interpreted as the
potential energy $U_l(\vec{r}_\mathrm{A})$---the vdW potential---from
which the force $\vec{F}_l(\vec{r}_\mathrm{A})$ can be obtained
($\bm{\nabla}_{\!\!\mathrm{A}}$ $\!\equiv$
$\!\bm{\nabla}_{\!\vec{r}_\mathrm{A}}$):
\begin{equation}
\label{eq38}
\Delta E_l=\Delta E_l^{(0)}+U_l(\vec{r}_\mathrm{A}),
\end{equation}
\begin{equation}
\label{eq38.1}
\vec{F}_l(\vec{r}_\mathrm{A})
=-\bm{\nabla}_{\!\!\mathrm{A}}U_l(\vec{r}_\mathrm{A}).
\end{equation}
Equation (\ref{eq38.1}) can be interpreted in several ways. So it can
be regarded as giving the force in the Newtonian equation of motion
for the center-of-mass coordinate, which is further evaluated within
the frame of quantum mechanics (cf., e.g., the analysis of quantum
reflection in Ref.~\cite{Jacobi02}) or, if possible, also within the
frame of classical mechanics (cf., e.g., Ref.~\cite{Raskin69}). In any
case the center-of-mass motion should be sufficiently slow, so that it
(approximately) decouples from the electronic motion in the spirit of
a Born-Oppenheimer approximation. Equation (\ref{eq38.1}) can also be
regarded as determining the force that must be compensated for in the
case where the center-of-mass coordinate may be considered as a given
(classical) parameter controlled externally. Since there is no need
here to distinguish between the possible interpretations, the operator
hat can be dropped.

The leading-order energy shift is given by the second-order term
\begin{align}
\label{eq39}
\Delta E_l &= -\frac{1}{\hbar}\sum_k\sum_{\lambda=e,m}\mathcal{P}
 \int_0^{\infty}\frac{\mathrm{d}\omega}{\omega_{kl}+\omega}
 \int\mathrm{d}^3r\nonumber\\
&\qquad\times\,\big|\langle l|\langle{\{0\}|
 -\hat{\vec{d}}\cdot\hat{\vec{E}}(\vec{r}}_\mathrm{A})
 |\{\vec{1}_\lambda(\vec{r},\omega)\}\rangle|k\rangle\big|^2
\end{align}
[$\mathcal{P}$, principal part;
$|\{\vec{1}_\lambda(\vec{r},\omega)\}\rangle$ $\!\equiv$
$\!\vec{f}_\lambda^\dagger(\vec{r},\omega)|\{0\}\rangle$]. We recall
definitions (\ref{eq20})--(\ref{eq22}) and make use of commutation
relations (\ref{eq15}) and (\ref{eq16}) as well as the relation
(\ref{eq14}), leading to
\begin{align}
\label{eq39.1}
U_l(\vec{r}_\mathrm{A}) &= -\frac{\mu_0}{\pi}\sum_k
 \mathcal{P}\int_0^\infty\!\!\mathrm{d}\omega\,
 \frac{\omega^2}{\omega_{kl}+\omega}\nonumber\\
&\qquad\times\,\vec{d}_{lk}\cdot\mathrm{Im}\,
 \tens{G}^{(1)}(\vec{r}_\mathrm{A},\vec{r}_\mathrm{A},\omega)
 \cdot\vec{d}_{kl}
\end{align}
($\vec{d}_{lk}$ $\!=$ $\!\langle l|\hat{\bf d}|k\rangle$). Note that
in agreement with with Eq.~(\ref{eq38}) we have dropped the 
$\vec{r}_A$-independent part of the energy shift by replacing
$\tens{G}(\vec{r}_\mathrm{A},\vec{r}_\mathrm{A},\omega)$
$\!\mapsto$
$\!\tens{G}^{(1)}(\vec{r}_\mathrm{A},\vec{r}_\mathrm{A},\omega)$,
where according to
\begin{equation}
\label{eq40}
\tens{G}(\vec{r},\vec{r}',\omega)
 =\tens{G}^{(0)}(\vec{r},\vec{r}',\omega)
 +\tens{G}^{(1)}(\vec{r},\vec{r}',\omega)
\end{equation}
the Green tensor has been decomposed into the (translationally
invariant) bulk part $\tens{G}^{(0)}(\vec{r},\vec{r}',\omega)$
corresponding to the vacuum region the atom is situated in
plus the scattering part $\tens{G}^{(1)}(\vec{r},\vec{r}',\omega)$
that accounts for the presence of the magnetodielectic bodies. 

Equation (\ref{eq39.1}) can be rewritten in a more convenient form by
transforming the integral along the real frequency axis into an
integral along the (positive) imaginary frequency axis with the aid of
property (\ref{eq12}) together with the well-known large-frequency
behaviour of the scattering Green tensor \cite{Ho03}. The result is
\begin{equation}
\label{eq41}
U_l(\vec{r}_\mathrm {A})
 =U_l^\mathrm{or}(\vec{r}_\mathrm{A})
 +U_l^\mathrm{r}(\vec{r}_\mathrm{A}),
\end{equation}
where
\begin{equation}
\label{eq42}
U_l^\mathrm{or}(\vec{r}_\mathrm{A})
 =\frac{\hbar\mu_0}{2\pi}
 \int_0^{\infty} \mathrm{d}u\,u^2 
 \mathrm{Tr}\bigl[\bm{\alpha}_l^{(0)}(iu)
 \cdot\tens{G}^{(1)}(\vec{r}_\mathrm{A},\vec{r}_\mathrm{A},iu)
 \bigr]
\end{equation}
is the off-resonant part of the potential,
\begin{equation}
\label{eq43}
U_l^\mathrm{r}(\vec{r}_\mathrm{A})
 =-\mu_0\sum_k \Theta(\omega_{lk})\omega_{lk}^2\,
 \vec{d}_{lk}\!\cdot\!\mathrm{Re}\bigl[
 \tens{G}^{(1)}(\vec{r}_\mathrm{A},\vec{r}_\mathrm{A},\omega_{lk})
 \bigr]\!\cdot\!\vec{d}_{kl}
\end{equation}
[$\Theta(z)$, unit step function] is the resonant part due the poles
at $\omega=\pm\omega_{lk}$ for $\omega_{lk}$ $\!>$ $\!0$, and
\begin{equation}
\label{eq44}
\bm{\alpha}_l^{(0)}(\omega)
 =\lim_{\epsilon\to 0}
 \frac{2}{\hbar}\sum_k 
 \frac{\omega_{kl} \vec{d}_{lk}
 \vec{d}_{kl}}
{\omega_{kl}^2-\omega^2-i\omega\epsilon}
\end{equation}
is the (lowest-order) atomic polarizability tensor. Note that the
resonant part of the vdW potential, which is absent if the atom is
prepared in its ground state, will in general dominate over the
off-resonant part for excited-state atoms. In particular for an atom
in a spherically symmetric state, e.g., the ground state,
Eqs.~(\ref{eq42}) and (\ref{eq43}) reduce to
\begin{align}
\label{eq45}
&U_l^\mathrm{or}(\vec{r}_\mathrm{A})
 = \frac{\hbar\mu_0}{2\pi}\int_0^\infty
 \mathrm{d}u\,u^2 \alpha_l^{(0)}(iu)\,
 \mathrm{Tr}\,
 \tens{G}^{(1)}(\vec{r}_\mathrm{A},\vec{r}_\mathrm{A},iu),\\
\label{eq46}
&U_l^\mathrm{r}(\vec{r}_\mathrm{A})
 = -\frac{\mu_0}{3}\sum_k\Theta(\omega_{lk})\omega_{lk}^2
 |\vec{d}_{lk}|^2
 \mathrm{Re}\bigl[\mathrm{Tr}
 \tens{G}^{(1)}(\vec{r}_\mathrm{A}\vec{r}_\mathrm{A},\omega_{lk})
 \bigr],
\end{align}
where
\begin{equation}
\label{eq47}
\alpha_l^{(0)}(\omega)
=\lim_{\epsilon\to 0}\frac{2}{3\hbar}\sum_k
\frac{\omega_{kl}|\vec{d}_{lk}|^2}
{\omega_{kl}^2-\omega^2-i\omega\epsilon}\,.
\end{equation}

Equations (\ref{eq41})--(\ref{eq44}) give the vdW potential of an atom
which is prepared in an energy eigenstate and situated near an
arbitrary arrangement of linear, iso\-tropic, dispersing, and
absorbing magnetodielectric bodies as a result of lowest-order QED
perturbation theory. Needless to say that they also apply to
left-handed materials \cite{Pendry99,Smith00,Veselago68}, for which
standard (normal-mode) quantization runs into difficulties. Moreover,
the derivation given can be regarded as a foundation of results
obtained on the basis of (semiphenomenological) linear response theory
\cite{Wylie84,Henkel02,Kryszewski92}. It should be pointed out that
the ground-state potential obtained from Eq.~(\ref{eq42}) can
equivalently be expressed in terms of an integral along the positive
(real) frequency axis, namely
\begin{align}
\label{eq48}
&U_0(\vec{r}_\mathrm{A}) = U_0^\mathrm{or}(\vec{r}_\mathrm{A})
= -\frac{\hbar\mu_0}{2\pi}
\nonumber\\&\quad\times\,
\int_0^\infty\mathrm{d}\omega\,\omega^2
\mathrm{Im}\,\bigl\{\mathrm{Tr}\,\bigl[
 \bm{\alpha}_0^{(0)}(\omega)\cdot
 \tens{G}^{(1)}(\vec{r}_\mathrm{A},\vec{r}_\mathrm{A},\omega)\bigr]
 \bigr\}.
\end{align}
This form allows for a simple physical interpretation of the vdW
potential as being due to the vacuum fluctuations of the electric
field inducing an electric dipole moment of the atom, together with 
the ground-state fluctuations of the atomic electric dipole moment
inducing an electric field \cite{Henkel02}.


\subsection{Dynamical theory}
\label{sec3.2}

A number of issues regarding the CP force can not be addressed within
the framework of (time-independent) perturbation theory. First, it is
known that the presence of macroscopic bodies can give rise to a
considerable change in the atomic level structure by inducing shifts
and broadenings of atomic transitions---an effect that is clearly
not accounted for in the lowest-order atomic polarizability as given
by Eq.~(\ref{eq44}). Second, spontaneous decay of an atom initially
prepared in an excited state will necessarily induce a dynamical
evolution of the force, a description of which is beyond the scope of
a time-independent theory. Third, the perturbative treatment does not
answer the question of the force acting on an atom not prepared in an
eigenstate of the atomic Hamiltonian (\ref{eq27}). Fourth, it seems
difficult to generalize the perturbative method towards a theory that
allows for electromagnetic fields prepared in arbitrary states---thus 
extending the concept of CP forces beyond a pure vacuum theory.
And fifth, perturbative methods break down completely in the case of
strong atom-field coupling.

In order to obtain an improved understanding of the CP force, we
consider a dynamical theory, the starting point being the Lorentz
force as appearing in the center-of-mass equation of motion of the
atom. Using Hamiltonian (\ref{eq33.1}) together with
Eqs.~(\ref{eq26}), (\ref{eq27}), and (\ref{eq31}) and recalling
definitions (\ref{eq29}) and (\ref{eq33}), one can verify that
\begin{equation}
\label{eq49}
m_\mathrm{A}\dot{\hat{\vec{r}}}_\mathrm{A}
 =\frac{i}{\hbar} \bigl[\hat{H},m_\mathrm{A} 
 \hat{\vec{r}}_\mathrm{A}\bigr]
 =\hat{\vec{p}}_\mathrm{A}
 +\hat{\vec{d}}\times\hat{\vec{B}}(\hat{\vec{r}}_\mathrm{A}),
\end{equation}
hence the total Lorentz force $\hat{\vec{F}}$ is given 
according to
\begin{align}
\label{eq50}
m_\mathrm{A}\ddot{\hat{\vec{r}}}_\mathrm{A} 
&= \hat{\vec{F}}=\frac{i}{\hbar} 
 \bigl[\hat{H},\hat{\vec{p}}_\mathrm{A}\bigr]
 +\frac{\mathrm{d}}{\mathrm{d}t}
 \bigl[\hat{\vec{d}}
 \times\hat{\vec{B}}(\hat{\vec{r}}_\mathrm{A})\bigr]
 \nonumber\\
&= \biggl\{\bm{\nabla}\bigl[
 \hat{\vec{d}}\cdot\hat{\vec{E}}(\vec{r})\bigr]
 +\frac{\mathrm{d}}{\mathrm{d}t}
 \bigl[\hat{\vec{d}}\times\hat{\vec{B}}(\vec{r})\bigr]
 \biggr\}_{\vec{r}=\hat{\vec{r}}_\mathrm{A}},
\end{align}
where (in the last step) magnetic dipole terms have been discarded in
consistency with the electric dipole approximation made, and a
nonrelativistic center-of-mass motion of the atom has been
assumed. Taking the expectation value with respect to the field
state and the internal atomic state yields an expression for the force
governing the center-of-mass motion,
\begin{equation}
\label{eq51}
\bigl\langle\hat{\vec{F}}\bigr\rangle
=\biggl\{\bm{\nabla}\bigl\langle
 \hat{\vec{d}}\cdot\hat{\vec{E}}(\vec{r})\bigr\rangle
 +\frac{\mathrm{d}}{\mathrm{d}t}
 \bigl\langle\hat{\vec{d}}
 \times\hat{\vec{B}}(\vec{r})\bigr\rangle
 \biggr\}_{\vec{r}=\hat{\vec{r}}_\mathrm{A}}.
\end{equation}
Equation (\ref{eq51}) together with Eqs.~(\ref{eq20})--(\ref{eq23}) and
Eq.~(\ref{eq32}) can be used to calculate the force in case of 
arbitrary (internal) atomic states, arbitrary field states, and
both weak and strong atom-field coupling. Obtaining an explicit
expression for the---in general time-dependent---force that only
depends on the initial conditions requires solving the atom-field
dynamics, i.e., $\hat{\vec{d}}$ $\!=$ $\!\hat{\vec{d}}(t)$,
$\hat{\vec{E}}(\vec{r})$ $\!=$ $\!\hat{\vec{E}}(\vec{r},t)$,
$\hat{\vec{B}}(\vec{r})$ $\!=$ $\!\hat{\vec{B}}(\vec{r},t)$, as
governed by Hamiltonian (\ref{eq33.1}) together with
Eqs.~(\ref{eq26}), (\ref{eq27}), and (\ref{eq31}).

In order to compare with the perturbative results of
Sec.~\ref{sec3.1}, we will calculate the force for the particular 
case of the body-assisted field being initially prepared in the
vacuum state $|\{0\}\rangle$ and the atom being initially prepared in
an energy eigenstate $|l\rangle$, so that the initial density
operator can be written as 
\begin{equation}
\label{eq52}
\hat{\varrho}=|\{0\}\rangle\langle\{0\}|\otimes|l\rangle\langle l|.
\end{equation}
We further assume that the atom-field coupling is weak, such that,
for chosen center-of-mass coordinate, the equations for the internal
atomic motion can be solved in the well-known Markov approximation. 
The physical meaning of the force determined in this way is 
basically the same as in the perturbative treatment, so
that---according to the remarks below Eq.~(\ref{eq38.1})---the
center-of-mass coordinate may be again regarded as being either 
a dynamical (operator-valued) variable or a ($c$-number) parameter.    
Therefore we will again drop the operator hat in what follows. 
In any case, the condition  
\begin{equation}
\label{eq52.1}
\tens{G} [\vec{r},\vec{r}_\mathrm{A}(t\!+\!\Delta t),\omega]\simeq
\tens{G} [\vec{r},\vec{r}_\mathrm{A}(t),\omega] \
\mathrm{for}\ \Delta t \le \Gamma_\mathrm{C}^{-1}
\end{equation}
must be satisfied in order to assure the validity of the
Born-Oppenheimer type approximation, where $\Gamma_\mathrm{C}$ is a
characteristic intra-atomic decay rate. For a non-degene\-rate system
Eq.~(\ref{eq51}) then leads to \cite{Buhmann04b}
\begin{equation}
\label{eq53}
\bigl\langle\hat{\vec{F}}(t)\bigr\rangle
=\sum_{m} \sigma_{mm}(t)\vec{F}_m(\vec{r}_\mathrm{A}),
\end{equation}
where 
\begin{align}
\label{eq54}
&\vec{F}_m(\vec{r}_\mathrm{A}) 
 =\frac{\mu_0}{2\pi} \sum_{k}
 \int_0^\infty\mathrm{d}\omega\,\omega^2
 \nonumber\\
&\quad\times\,\frac{\bm{\nabla}_\mathrm{A}
 \vec{d}_{mk}\!\cdot\!\mathrm{Im}\bigl[
 \tens{G}^{(1)}(\vec{r}_\mathrm{A},\vec{r}_\mathrm{A},\omega) 
 \bigr]\!\cdot\vec{d}_{km}}
 {\omega\!+\!\tilde{\omega}_{km}(\vec{r}_\mathrm{A})
 \!-\!i[\Gamma_k(\vec{r}_\mathrm{A})
 \!+\!\Gamma_m(\vec{r}_\mathrm{A})]/2}
 + \mathrm{H.c.}\,,
\end{align}
and the internal atomic density matrix elements
$\sigma_{mm}(t)$ obey the balance equations
\begin{equation}
\label{eq55}
\dot{\sigma}_{mm}(t)
 = -\Gamma_m(\vec{r}_\mathrm{A})\sigma_{mm}(t)
 +\sum_n\Gamma_n^m(\vec{r}_\mathrm{A})\sigma_{nn}(t)
\end{equation}
together with the initial condition $\sigma_{mm}(0)$ $\!=$
$\!\delta_{ml}$. In Eqs. (\ref{eq54}) and (\ref{eq55}), 
\begin{eqnarray}
\label{eq56}
\tilde{\omega}_{mn}(\vec{r}_\mathrm{A})
 =\omega_{mn}+\delta\omega_m(\vec{r}_\mathrm{A})
 -\delta\omega_n(\vec{r}_\mathrm{A})
\end{eqnarray}
are the body-induced, position-dependent, shifted atomic transition
frequencies, where
\begin{equation}
\label{eq57}
\delta\omega_m(\vec{r}_\mathrm{A})
 =\sum_k\delta\omega_m^k(\vec{r}_\mathrm{A})
\end{equation}
with
\begin{equation}
\label{eq58}
\delta\omega_m^k(\vec{r}_\mathrm{A})
 =\frac{\mu_0}{\pi\hbar}
 {\cal P}\!\int_0^\infty \!\!\mathrm{d}\omega\,\omega^2
 \frac{\vec{d}_{mk}\!\cdot\!\mathrm{Im}\bigl[
 \tens{G}^{(1)}(\vec{r}_\mathrm{A},\vec{r}_\mathrm{A},\omega)
 \bigr]\!\cdot\vec{d}_{km}}
 {\tilde{\omega}_{mk}(\vec{r}_\mathrm{A})-\omega},
\end{equation}
and
\begin{equation}
\label{eq59}
\Gamma_m(\vec{r}_\mathrm{A})
 =\sum_k\Gamma_m^k(\vec{r}_\mathrm{A})
\end{equation}
are the position-dependent level widths, where
\begin{align}
\label{eq60}
\Gamma_m^k(\vec{r}_\mathrm{A})
&= \frac{2\mu_0 }{\hbar}\,
 \Theta[\tilde{\omega}_{mk}(\vec{r}_\mathrm{A})]
 [\tilde{\omega}_{mk}(\vec{r}_\mathrm{A})]^2\nonumber\\
&\quad\times\vec{d}_{mk}\cdot\mathrm{Im}\bigl\{
 \tens{G}[\vec{r}_\mathrm{A},\vec{r}_\mathrm{A},
 \tilde{\omega}_{mk}(\vec{r}_\mathrm{A})]\bigr\}
 \cdot\vec{d}_{km}.
\end{align}
Note that Eqs.~(\ref{eq56})--(\ref{eq58}) have to be solved
self-consistent\-ly, where the position-independent Lamb-shift
terms resulting from
$\tens{G}^{(0)}\,(\vec{r}_\mathrm{A},\vec{r}_\mathrm{A},\omega)$
[recall Eq.~(\ref{eq40})] have been absorbed in the transition
frequencies $\omega_{mn}$. 

In a similar way as in Sec.~\ref{sec3.1} [cf. the remark above
Eq.~(\ref{eq41})], Eq.~(\ref{eq54}) can be simplified by means of
contour integral techniques, resulting in
\begin{equation}
\label{eq61}
\vec{F}_m(\vec{r}_\mathrm{A})
=\vec{F}_m^\mathrm{or}(\vec{r}_\mathrm{A})
+\vec{F}_m^\mathrm{r}(\vec{r}_\mathrm{A}),
\end{equation}
where
\begin{align}
\label{eq62}
&\vec{F}_m^\mathrm{or}(\vec{r}_\mathrm{A})
 =-\frac{\hbar\mu_0}{4\pi}\int_0^\infty\mathrm{d}u u^2
 \big[(\alpha_m)_{ij}(\vec{r}_\mathrm{A},iu)
 \nonumber\\
&\qquad +(\alpha_m)_{ij}(\vec{r}_\mathrm{A},-iu)\big]
 \bm{\nabla}_{\!\!\mathrm{A}}
 G^{(1)}_{ij}(\vec{r}_\mathrm{A},\vec{r}_\mathrm{A},iu)
\end{align}
and
\begin{align}
\label{eq63}
&{\vec{F}}_m^\mathrm{r}(\vec{r}_\mathrm{A})
 =\frac{\mu_0}{2}\sum_{k}
 \Theta({\tilde{\omega}_{mk}})\Omega_{mk}^2(\vec{r}_\mathrm{A})
\nonumber\\&\quad\times\,
 \Bigl\{\bm{\nabla}\vec{d}_{mk}\cdot
 \tens{G}^{(1)}[\vec{r},\vec{r},\Omega_{mk}(\vec{r}_\mathrm{A})]
 \cdot\vec{d}_{km}\Bigr\}_{\vec{r}=\vec{r}_\mathrm{A}}
 +\mathrm{H.c.}\,,
\end{align}
with
\begin{align}
\label{eq64}
&\bm{\alpha}_{m}(\vec{r}_\mathrm{A},\omega) 
 =\frac{1}{\hbar}\sum_k\biggl\{
 \frac{\vec{d}_{mk}\vec{d}_{km}}
 {\tilde{\omega}_{km}(\vec{r}_\mathrm{A})
 \!-\!\omega\!-\!i[\Gamma_k(\vec{r}_\mathrm{A})
 \!+\!\Gamma_m(\vec{r}_\mathrm{A})]/2}
\nonumber\\&\qquad 
 +\frac{\vec{d}_{km}\vec{d}_{mk}}
 {\tilde{\omega}_{km}(\vec{r}_\mathrm{A}) 
 \!+\!\omega\!+\!i[\Gamma_k(\vec{r}_\mathrm{A})
 \!+\!\Gamma_m(\vec{r}_\mathrm{A})]/2}\biggr\},
\end{align}
being the (exact) body-assisted atomic polarizability and
\begin{equation}
\label{eq65}
\Omega_{mk}(\vec{r}_\mathrm{A})
 =\tilde{\omega}_{mk}(\vec{r}_\mathrm{A})
 +i[\Gamma_m(\vec{r}_\mathrm{A})
 +\Gamma_k(\vec{r}_\mathrm{A})]/2,
\end{equation}
denoting the shifted and broadened atomic transition frequencies.  

The dynamical result differs from the perturbative one
in several respects. {F}rom Eqs.~(\ref{eq53}) and (\ref{eq55}) 
it is seen that---as expected---spontaneous decay gives rise to a
temporal evolution of the CP force, which is governed by the temporal
evolution of the respective diagonal density matrix elements. 
Only if the atom is initially (at time $t$ $\!=$ $\!0$) prepared in
its ground state ($l$ $\!=$ $\!0$), a time-indepent force
\begin{equation}
\label{eq66}
\bigl\langle\hat{\vec{F}}(t)\bigr\rangle
 =\bigl\langle\hat{\vec{F}}(0)\bigr\rangle
 =\vec{F}_0(\vec{r}_\mathrm{A})
 =\vec{F}_0^\mathrm{or}(\vec{r}_\mathrm{A})
\end{equation}
can be observed. 
When on the contrary the atom is initially prepared in an excited state
($l$ $\!\neq$ $\!0$), then the initial single-component force
\begin{equation}
\label{eq67-1}
\bigl\langle\hat{\vec{F}}(0)\bigr\rangle
 = \vec{F}_l(\vec{r}_\mathrm{A})
\end{equation}
can be observed only for times 
$t$ $\!\ll$ $\!\Gamma_l^{-1}(\vec{r}_\mathrm{A})$, i.e.,
\begin{equation}
\label{eq67}
\bigl\langle\hat{\vec{F}}(t)\bigr\rangle
 \simeq\vec{F}_l(\vec{r}_\mathrm{A}),\quad
 t\ll\Gamma_l^{-1}(\vec{r}_\mathrm{A}).
\end{equation}
In the further course of time the single-component force evolves into
a multi-component force at intermediate times (the atom being in a
mixed state) and eventually reduces to the ground-state force for
large times,
\begin{equation}
\label{eq68}
\bigl\langle\hat{\vec{F}}(t)\bigr\rangle
 \simeq\vec{F}_0(\vec{r}_\mathrm{A}),\quad
 t\gg\Gamma_m^{-1}(\vec{r}_\mathrm{A})
 \ \forall\ m\le l.
\end{equation}

\begin{figure}[!t!]
\noindent
\begin{center}
\includegraphics[width=\linewidth]{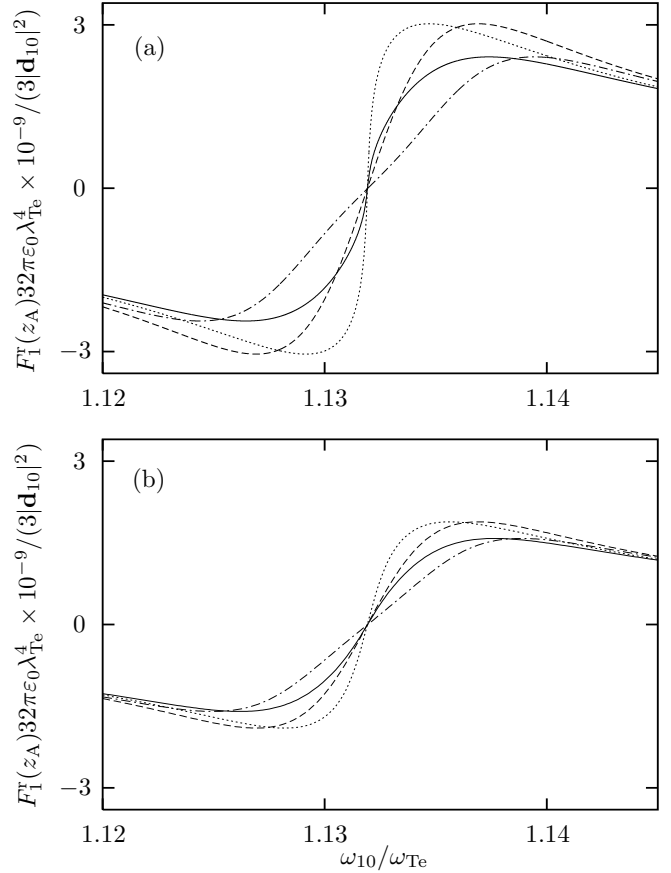}
\end{center}
\caption{
\label{fig0}
Resonant part of the CP force on an excited two-level atom that is
situated at distance 
(a) $z_\mathrm{A}/\lambda_\mathrm{Te}$ $\!=$ $\!0.0075$ and
(b) $z_\mathrm{A}/\lambda_\mathrm{Te}$ $\!=$ $\!0.009$
($\lambda_\mathrm{Te}$ $\!=$ $\!2\pi c/\omega_\mathrm{Te}$)
of a semi-infinite dielectric half space and whose transition dipole
moment is perpendicular to the interface, as a function of the atomic
transition frequency (solid lines), where
$\omega_\mathrm{Pe}/\omega_\mathrm{Te}$ $\!=$ $\!0.75$,
$\gamma_\mathrm{e}/\omega_\mathrm{Te}$ $\!=$ $\!0.01$,
$\omega_\mathrm{Te}^2|\vec{d}_{10}|^2/(3\pi\hbar\varepsilon_0c^3)$
$\!=$ $\!10^{-7}$. For comparison, both the perturbative result, i.e.,
$\delta\omega_m(z_\mathrm{A})$ $\!=$ $\!\Gamma_m(z_\mathrm{A})$ $\!=$
$\!0$ (dashed lines) and the results obtainable by only considering
the effect of level shifting, i.e., $\Gamma_m(z_\mathrm{A})$ $\!=$
$\!0$ (dotted lines) or only considering the effect of level
broadening, i.e., $\delta\omega_m(z_\mathrm{A})$ $\!=$ $\!0$
(dash-dotted lines) are shown.
}
\end{figure}

Thus the perturbative treatment of Sec.~\ref{sec3.1} effectively turns
out to be an approximate calculation of the force components 
$\vec{F}_l(\vec{r}_\mathrm{A})$, thereby disregarding the effects of
level shifting and broadening. On the contrary, the force components
as given by Eqs.~(\ref{eq62}) and (\ref{eq63}) depend on the correct
shifted and broadened atomic transition frequencies (\ref{eq65}) that
are observed in the presence of the bodies, and hence also on the
correct body-assisted position-dependent polarizability (\ref{eq64}).
Inspection of Eqs.~(\ref{eq62})--(\ref{eq65}) reveals that the
frequency shifts affect both ground- and excited-state force
components, whereas the decay-induced level broadening only has a
noticeable (reducing) effect on the resonant force components present
for atoms in excited states. For example, the resonant force component
$\vec{F}_1^\mathrm{r}(z_\mathrm{A})$ $\!=$
$\!F_1^\mathrm{r}(z_\mathrm{A})\vec{e}_z$ acting on an excited
two-level atom situated at a very small distance $z_\mathrm{A}$ from
a semi-infinite dielectric half-space is given by \cite{Buhmann04b}
\begin{equation}
\label{eq68.1}
F_1^\mathrm{r}(z_\mathrm{A})
=-\frac{3|\vec{d}_{10}|^2}{32\pi\varepsilon_0z_\mathrm{A}^4}
\frac{|\varepsilon[\Omega_{10}(z_\mathrm{A})]|^2-1}
{|\varepsilon[\Omega_{10}(z_\mathrm{A})]+1|^2}\,.
\end{equation}
{F}rom Fig.~\ref{fig0}, which shows $F_1^\mathrm{r}(z_\mathrm{A})$ 
for the case of the permittivity being modelled by 
\begin{equation}
\label{eq84}
\varepsilon(\omega) = 1+\,\frac{\omega_\mathrm{Pe}^2}
{\omega_\mathrm{Te}^2-\omega^2-i\omega\gamma_\mathrm{e}}\,,
\end{equation}
it is seen that the typical dispersion profile already observed in the
perturbative treatment becomes narrower due to the level shifting
while the level broadening has the effect of lowering and broadening
the dispersion profile. The different behaviour of the resonant and
off-resonant force components with respect to the effect of level
broadening is closely related to the fact that  
$\vec{F}_m^\mathrm{r}(\vec{r}_\mathrm{A})$ [Eq.~(\ref{eq63}) together
with Eq.~(\ref{eq65})] is linear in $\Gamma_m(\vec{r}_\mathrm{A})$ in
lowest order, whereas $\vec{F}_m^\mathrm{or}(\vec{r}_\mathrm{A})$
[Eq.~(\ref{eq62}) together with Eq.~(\ref{eq64})] is only quadratic in
$\Gamma_m(\vec{r}_\mathrm{A})$, as a Taylor expansion shows.
Physically, this can be understood from the argument that the off-resonant
force components can be regarded as being due to virtual transitions 
which happen on very short time scales, so that spontaneous decay
cannot have a major influence. 

It is worth noting that the additional position-depen\-dence
introduced via the frequency shifts and broadenings has the effect
that even the ground-state force cannot be derived, in general, from a
potential in the way prescribed by Eqs.~(\ref{eq38}) and
(\ref{eq38.1}) in Sec.~\ref{sec3.1}. While the force as given by
Eq.~(\ref{eq53}) can of course still be written as a (time-dependent)
potential force provided that the force components as given by
Eqs.~(\ref{eq61})--(\ref{eq63}) are irrotational vectors (which is
indeed the case for, e.g., an atom in the presence of planarly,
spherically or cylindrically multilayered media), there may be
situations where this is not possible, implying that
Eqs.~(\ref{eq61})--(\ref{eq63}) can not be derived from an energy
expression in the way given by Eqs.~(\ref{eq38}) and (\ref{eq38.1}) in
principle. 

Clearly, the above mentioned effects of level shifting and broadening
can only become relevant when the atom is situated sufficiently close
to a body surface. As already mentioned, when the frequency shifts and
broadenings can be neglected, $\delta\omega_m(\vec{r}_\mathrm{A})$
$\!\to$ $\!0$, $\Gamma_m(\vec{r}_\mathrm{A})$ $\!\to$ $\!0$, 
then the dynamical result for the force components
$\vec{F}_m(\vec{r}_\mathrm{A})$ calculated by using Eq.~(\ref{eq61})
together with Eqs.~(\ref{eq62})--(\ref{eq64}) simplifies to the
perturbative one calculated from Eq.~(\ref{eq38.1}) together with 
Eqs. (\ref{eq41})--(\ref{eq44}). If necessary, the level shifts 
could of course be easily introduced in the perturbative formulas by
replacing the bare transition frequencies with the shifted ones
[$\omega_{mn}$ $\!\mapsto$
$\!\tilde{\omega}_{mn}(\vec{r}_\mathrm{A})$]. On the contrary,
introduction of the level broadening is not so straightforward. In
particular, the results of the dynamical theory can not be reproduced
from the perturbative results by making the replacement
$\bm{\alpha}_0^{(0)}(\omega)$
$\!\mapsto$~$\!\bm{\alpha}_0(\vec{r}_\mathrm{A},\omega)$ in the
off-resonant force components (as done, e.g., in
Ref.~\cite{Kryszewski92}) and replacing the bare transition
frequencies by complex ones according to $\omega_{mn}$ $\!\mapsto$
$\!\Omega_{mn}(\vec{r}_A)$ in the resonant components. Hence, the
perturbative results as given in Sec.~\ref{sec3.1} may be regarded as
a reasonable approximation only for the ground-state CP force, which
is solely determined by the off-resonant component
$\vec{F}_0^\mathrm{or}(\mathbf{r}_A)$ and thus effectively not
influenced by level broadening.  


\section{Ground-state atom within magnetodielectric multilayer system}
\label{sec4}

To study the competing effects of electric and magnetic properties
of the bodies on the CP force, let us consider a ground-state atom
placed within a magnetodielectric multilayer system. {F}rom the
arguments given above, we base, for simplicity, the calculations on
the perturbative analysis, calculating the ground-state vdW potential
$U_0(\vec{r}_A)$ $\!=$~$\!U_0^\mathrm{or}(\vec{r}_A)$
according to Eq.~(\ref{eq45}) together with Eq.~(\ref{eq47}).
\begin{figure}[!t!]
\noindent
\begin{center}
\includegraphics[width=\linewidth]{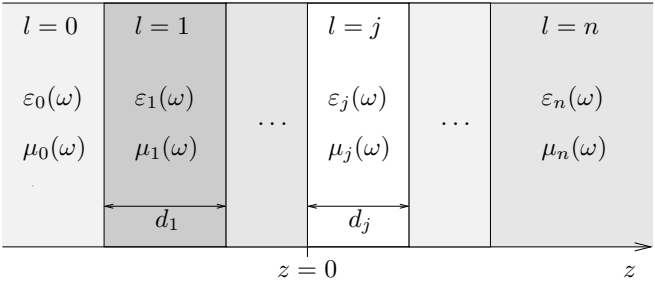}
\end{center}
\caption{
\label{fig1}
Sketch of the planar multilayer system.}
\end{figure}
 
The planar multilayer system can be characterized as a stack of $n$ 
$\!+$ $\!1$ layers labelled by $l$ ($l$ $\!=$ $0,\ldots,n$) of
thicknesses $d_l$ with planar parallel boundary surfaces, where
$\varepsilon(\vec{r},\omega)$ $\!=$ $\!\varepsilon_l(\omega)$ and
$\mu(\vec{r},\omega)$ $\!=$ $\mu_l(\omega)$. The coordinate system
is chosen such that the layers are perpendicular to the $z$ axis and 
extend from \mbox{$z$ $\!=$ $\!0$} to $z$ $\!=$ $\!d_l$ for $l$
$\!\neq$ $\!0,n$ and from \mbox{$z$ $\!=$ $\!0$} to $z$ $\!=$
$\!-\infty$ ($\infty$) for $l$ $\!=$ $\!0$ ($n$) (cf. Fig.~\ref{fig1},
where the position $z$ $\!=$ $\!0$ refers to layer $j$). The
scattering part of the Green tensor at imaginary frequencies for
$\vec{r}$ and $\vec{r}'$ in layer $j$ can be given by \cite{Chew95}
\begin{equation}
\label{eq69}
\tens{G}^{(1)}(\vec{r},\vec{r}',iu) =
\int\mathrm{d}^2q\,
e^{i\vec{q}\cdot(\vec{r}-\vec{r}')}
\tens{G}^{(1)}(\vec{q},z,z',iu)
\end{equation}
($\mathbf{q}\perp\mathbf{e}_z$). Here,
\begin{align}
\label{eq70}
&\tens{G}^{(1)}(\vec{q},z,z',iu) = \frac{\mu_j(iu)}{8\pi^2b_j}
 \sum_{\sigma=s,p}\biggl\{ \frac{r^\sigma_{j-}r^\sigma_{j+}
 e^{-2b_jd_j}}{D_j^\sigma}\nonumber\\
&\quad\times\,\Bigl[\vec{e}_\sigma^+\vec{e}_\sigma^+ 
 e^{-b_j(z-z')}
 +\vec{e}_\sigma^-\vec{e}_\sigma^-
 e^{b_j(z-z')}\Bigr]\nonumber\\
&\qquad+\frac{1}{D_j^\sigma}
 \Bigl[\vec{e}_\sigma^+\vec{e}_\sigma^- r^\sigma_{j-} 
 e^{-b_j(z+z')}\nonumber\\
&\qquad\qquad\quad+\vec{e}_\sigma^-\vec{e}_\sigma^+
r^\sigma_{j+}
 e^{-2b_jd_j}e^{b_j(z+z')}\Bigr]\biggr\}
\end{align}
for $j$ $\!>$ $\!0$, where
\begin{equation}
\label{eq71}
\vec{e}_s^\pm=\vec{e}_q\times\vec{e}_z,
 \quad\vec{e}_p^\pm=-\frac{1}{k_j}(iq\vec{e}_z
 \pm b_j\vec{e}_q)
\end{equation}
($\vec{e}_q$ $\!=$ $\!\vec{q}/q$, $q$ $\!=$ $\!|\vec{q}|$) with
\begin{equation}
\label{eq72}
k_j=\frac{u}{c}\sqrt{\varepsilon_j(iu)\mu_j(iu)}
\end{equation}
are the polarization vectors for $s$- and $p$-polarized waves
propagating in the positive ($+$) and negative ($-$) $z$-direc\-tions,
$r^\sigma_{j-}$ and $r^\sigma_{j+}$ are the generalized coefficients
for reflection at the left/right boundary of layer $j$, which can be
calculated with the aid of the recursive relations
\begin{alignat}{1}
\label{eq73}
&r^s_{l\pm}=
\frac{\left(\frac{\mu_{l\pm 1}}{b_{l\pm 1}}-\frac{\mu_l}{b_l}\right)
+\left(\frac{\mu_{l\pm 1}}{b_{l\pm 1}}+\frac{\mu_l}{b_l}\right)
e^{-2b_{l\pm 1}d_{l\pm 1}}r^s_{l\pm 1\pm}}
{\left(\frac{\mu_{l\pm 1}}{b_{l\pm 1}}+\frac{\mu_l}{b_l}\right)
+\left(\frac{\mu_{l\pm 1}}{b_{l\pm 1}}-\frac{\mu_l}{b_l}\right)
e^{-2b_{l\pm 1}d_{l\pm 1}}r^s_{l\pm 1\pm}}\, ,\\
\label{eq74}
&r^p_{l\pm}
=\frac{\left(\frac{\varepsilon_{l\pm 1}}{b_{l\pm 1}}
-\frac{\varepsilon_l}{b_l}\right)
+\left(\frac{\varepsilon_{l\pm 1}}{b_{l\pm 1}}
+\frac{\varepsilon_l}{b_l}\right)
e^{-2b_{l\pm 1}d_{l\pm 1}}r^p_{l\pm 1\pm}}
{\left(\frac{\varepsilon_{l\pm 1}}{b_{l\pm 1}}
+\frac{\varepsilon_l}{b_l}\right)
+\left(\frac{\varepsilon_{l\pm 1}}{b_{l\pm 1}}
-\frac{\varepsilon_l}{b_l}\right)
e^{-2b_{l\pm 1}d_{l\pm 1}}r^p_{l\pm 1\pm}}
\end{alignat}
($l$ $\!=$ $\!1,\ldots,j$ for $r^\sigma_{l-}$, 
$l$ $\!=$ $\!j,\ldots,n$ $\!-$ $\!1$ for $r^\sigma_{l+}$,
\mbox{$r^\sigma_{0-}$ $\!=$ $\!r^\sigma_{n+}$ $\!=$ $\!0$}),
\begin{equation}
\label{eq75}
b_l = \sqrt{\frac{u^2}{c^2}\ \varepsilon_l(iu)\mu_l(iu) + q^2}
\end{equation}
is the imaginary part of the $z$-component of the wave vector in layer
$l$,
and
\begin{equation}
\label{eq76}
D_j^\sigma=1-r_{j-}^\sigma r_{j+}^\sigma e^{-2b_jd_j}.
\end{equation}

Let the atom be situated in the otherwise empty layer $j$, i.e., 
$\varepsilon_j(iu)$ $\!=$ $\!\mu_j(iu)$ $\!\equiv$ $\!1$ and
\begin{equation}
\label{eq77}
b_j=\sqrt{\frac{u^2}{c^2}+q^2} \equiv b.
\end{equation}
To calculate the vdW potential, we substitute Eq.~(\ref{eq69})
together with Eq.~(\ref{eq70}) into Eq.~(\ref{eq45}), thereby omitting
irrelevant position-independent terms [recall that
$U_0^\mathrm{or}(\vec{r}_A)$ $\!=$~$\!U_0(\vec{r}_A)$].
Evaluating the trace with the aid of the relations
\begin{align}
\label{eq78}
&\vec{e}_s^\pm\cdot\vec{e}_s^\pm
=\vec{e}_s^\pm\cdot\vec{e}_s^\mp=1,\\
\label{eq79}
&\vec{e}_p^\pm\cdot\vec{e}_p^\pm=1, 
\quad \vec{e}_p^\pm\cdot\vec{e}_p^\mp
=-1-2\left(\frac{qc}{u}\right)^2,
\end{align}
which directly follow from Eqs.~(\ref{eq71}), (\ref{eq72}), and
(\ref{eq77}), we realize that the resulting integrand of the
$\vec{q}$-integral only depends on $q$. Thus after introducing
polar coordinates in the $q_xq_y$-plane, we can easily perform the
angular integration, leading to
\begin{align}
\label{eq80}
&U_0(z_\mathrm{A}) = \frac{\hbar\mu_0}{8\pi^2}
 \int_0^{\infty} \mathrm{d} u \,u^2 \alpha_0^{(0)}(iu)
 \int_0^\infty\mathrm{d}q\,\frac{q}{b}\nonumber\\
&\quad\times\,\Biggl\{
 e^{-2bz_\mathrm{A}}\biggl[\frac{r_{j-}^s}{D_j^s}
 -\biggl(1+2\frac{q^2c^2}{u^2}\biggr)
 \frac{r_{j-}^p}{D_j^p}\biggr]\nonumber\\
&\qquad\quad+e^{-2b(d_j-z_\mathrm{A})}
 \biggl[\frac{r_{j+}^s}{D_j^s}-\biggl(1+2\frac{q^2c^2}{u^2}\biggr)
 \frac{r_{j+}^p}{D_j^p}\biggr]\Biggr\}.
\end{align}
Note that Eq.~(\ref{eq70}) and thus Eq.~(\ref{eq80}) also apply to
the case $j$ $\!=$ $\!0$ if $d_0$ is formally set equal to zero
\mbox{($d_0$ $\!\equiv$ $\!0$)}.

Equation (\ref{eq80}) together with Eq.~(\ref{eq47}) and
Eqs.~(\ref{eq73})--(\ref{eq77}) gives the vdW potential of a
ground-state atom with\-in a general planar magnetodielectric
multilayer system in terms of the atomic polarizability and the
generalized reflection coefficients. Note that instead of calculating
these coefficients from the permittivities and permeabilities of the
individual layers via Eqs.~(\ref{eq73})--(\ref{eq75}) (as we shall do
in this paper), it is possible to determine them experimentally by
appropriate reflectivity measurements (cf., e.g.,
Ref.~\cite{Thakur04}). The coefficients $D_j^\sigma$
[Eq.~(\ref{eq78})] describe the effect of multiple reflections of
radiation at the two boundaries of the vacuum layer $j$ the atom is
placed in, as can be seen by expanding $D_j^\sigma$ according to 
\begin{equation}
\label{eq80.1}
\frac{1}{D_j^\sigma}=\sum_{n=0}^\infty \big(r_{j-}^\sigma e^{-b_jd_j}
 r_{j+}^\sigma e^{-b_jd_j}\big)^n\, .
\end{equation}
Multiple reflections within layer $j$ do obviously not occur if the
atom is placed in one of the semi-infinite outer layers ($j$
$\!=$~$\!n$), so
that Eq.~(\ref{eq80}) reduces to
\begin{align}
\label{eq81}
U_0(z_\mathrm{A}) 
& = \frac{\hbar\mu_0}{8\pi^2}
 \int_0^{\infty} \mathrm{d} u \,u^2 \alpha_0^{(0)}(iu)
 \int_0^\infty\mathrm{d}q\,
 \frac{q}{b}e^{-2bz_\mathrm{A}}\\\nonumber
&\qquad\qquad\times\,\biggl[r_{n-}^s
-\biggl(1+2\frac{q^2c^2}{u^2}\biggr)r_{n-}^p\biggr].
\end{align}


\subsection{Infinitely thick plate}
\label{sec4.1}

Let us apply Eqs.~(\ref{eq80}) and (\ref{eq81}) to some simple
systems and begin with an atom in front of a sufficiently thick
magnetodielectric plate which can be effectively regarded as a
semi-infinite half space [$n$ $\!=$ $\!j$ $\!=$ $\!1$, 
$\varepsilon_1(\omega)$ $\!=$ $\!\mu_1(\omega)$ $\!\equiv$ $\!1$,
$\varepsilon_0(\omega)$ $\!\equiv$ $\!\varepsilon(\omega)$,
$\mu_0(\omega)$ $\!\equiv$ $\!\mu(\omega)$]. Using Eqs.~(\ref{eq73})
and (\ref{eq74}) we find that the reflection coefficients in
Eq.~(\ref{eq81}) read ($b_0$ $\!\equiv$~$\!b_\mathrm{M}$)
\begin{eqnarray}
\label{eq82}
r_{n-}^s&=&
 \frac{\mu(iu)b-b_\mathrm{M}}{\mu(iu)b+b_\mathrm{M}}\,,\\
\label{eq83} 
r_{n-}^p&=& 
 \frac{\varepsilon(iu)b-b_\mathrm{M}}{\varepsilon(iu)b+b_\mathrm{M}}
 \,.
\end{eqnarray}
Note that Eq.~(\ref{eq81})  together with Eqs.~(\ref{eq82}) and
(\ref{eq83}) is equivalent to the result derived in
Ref.~\cite{Kryszewski92} semiclassically within the frame of linear
response theory. 

To further analyze Eqs.~(\ref{eq81})--(\ref{eq83}), let us model the
permittivity by Eq.~(\ref{eq84}) and the permeability by
\begin{equation}
\label{eq85}
\mu(\omega) = 1 +\,\frac{\omega_\mathrm{Pm}^2}
{\omega_\mathrm{Tm}^2-\omega^2-i\omega\gamma_\mathrm{m}}\,.
\end{equation} 
In the long-distance (retarded) limit, i.e., $z_\mathrm{A}$ $\!\gg\!$
$c/\omega_\mathrm{A}^-$, $z_\mathrm{A}$
$\!\gg$~$\!c/\omega_\mathrm{M}^-$ [$\omega_\mathrm{A}^-$ $\!=$
$\mathrm{min}(\{\omega_{k0}|k$ $\!=$ $\!1,2,\ldots\})$,
$\omega_\mathrm{M}^-$ $\!=$
$\mathrm{min}(\omega_\mathrm{Te},$ $\!\omega_\mathrm{Tm})$], 
Eqs.~(\ref{eq81})--(\ref{eq83}) reduce to (see Appendix \ref{appB})
\begin{equation}
\label{eq86}
U_0(z_\mathrm{A})=\frac{C_4}{z_\mathrm{A}^4}\,,
\end{equation}
where
\begin{align}
\label{eq87}
C_4
=& -\frac{3\hbar c\alpha_0^{(0)}(0)}{64\pi^2\varepsilon_0}
 \int_{1}^\infty\mathrm{d} v\, 
 \left[\left(\frac{2}{v^2}-\frac{1}{v^4}\right)\right.\nonumber\\
&\quad\times\,\frac{\varepsilon(0)v-\sqrt{\varepsilon(0)\mu(0)-1+v^2}}
 {\varepsilon(0)v+\sqrt{\varepsilon(0)\mu(0)- 1 + v^2 }}
 \nonumber\\
&\quad-\frac{1}{v^4}\,\left.
 \frac{\mu(0)v-\sqrt{\varepsilon(0)\mu(0)-1 + v^2}}
 {\mu(0)v+\sqrt{\varepsilon(0)\mu(0)-1 + v^2 }}\right],
\end{align}
while in the short-distance (nonretarded) limit, i.e.,
$z_\mathrm{A}$ $\!\ll$~$\!c/\omega_\mathrm{A}^+$ and/or
$z_\mathrm{A}$ $\!\ll$ $\!c/\omega_\mathrm{M}^+$ 
[$\omega_\mathrm{A}^+$ $\!=$ $\!\mathrm{max}(\{\omega_{k0}|k$
$\!=$ $\!1,2,\ldots\})$,
$\omega_\mathrm{M}^+$ $\!=\!$ 
$\mathrm{max}(\omega_\mathrm{Te},$ $\!\omega_\mathrm{Tm})$],
Eqs.~(\ref{eq81})--(\ref{eq83}) lead to (see Appendix \ref{appB})
\begin{equation}
\label{eq88}
U_0(z_\mathrm{A})= -\ \frac{C_3}{z_\mathrm{A}^3}
 +\frac{C_1}{z_\mathrm{A}}\,,
\end{equation}
where
\begin{equation}
\label{eq89}
C_3
 = \frac{\hbar}{16\pi^2\varepsilon_0}
 \int_0^\infty\mathrm{d}u\ \alpha_0^{(0)}(iu)
 \frac{\varepsilon(iu)-1}{\varepsilon(iu)+1}\ge 0
\end{equation}
and
\begin{align}
\label{eq90}
C_1 =\ & \frac{\mu_0\hbar}{16\pi^2}
 \int_0^\infty\mathrm{d}u\ u^2\alpha_0^{(0)}(iu)
 \biggl\{\frac{\varepsilon(iu)-1}{\varepsilon(iu)+1}\nonumber\\
&+ \frac{\mu(iu)-1}{\mu(iu)+1}
 + \frac{2\varepsilon(iu)[\varepsilon(iu)\mu(iu)-1]}
 {[\varepsilon(iu)+1]^2}
 \biggr\}\ge 0. 
\end{align}
It should be pointed out that this asymptotic behaviour 
also remains valid for multiresonance permittivities and
permeabilities of Drude-Lorentz type. Clearly, in this case
the minimum $\omega_\mathrm{M}^-$ and the maximum
$\omega_\mathrm{M}^+$ are defined with respect to all matter
resonances. 

\begin{figure}[!t!]
\noindent
\begin{center}
\includegraphics[width=\linewidth]{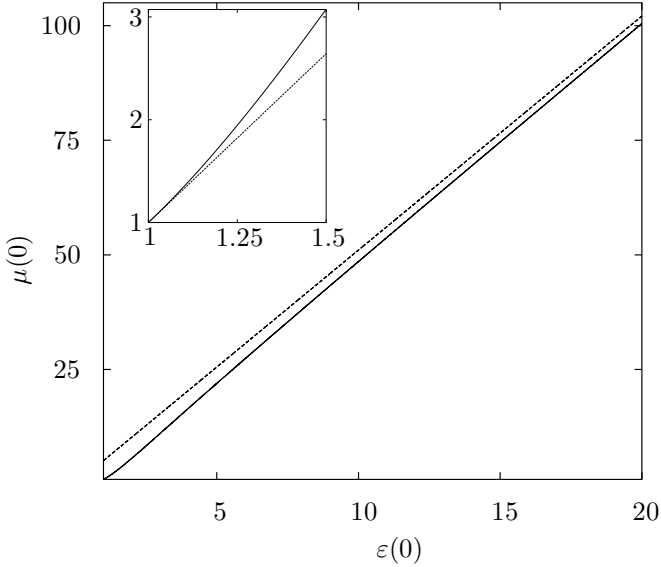}
\end{center}
\caption{
\label{fig5}
Border between attractive and repulsive long-distance vdW potentials
of an atom in front of an infinitely thick magnetodielectric plate
according to Eq.~(\ref{eq87}) ($C_4$ $\!=$ $\!0$). The broken curves
show the asymptotic behaviour as given by Eqs.~(\ref{eq94}) (inset)
and (\ref{eq96}). 
}
\end{figure}

Inspection of Eq.~(\ref{eq87}) reveals that the coefficient $C_4$ 
in Eq.~(\ref{eq86}) for the long-distance behaviour of the vdW
potential is negative (positive) for a purely electric 
(magnetic) plate, corresponding to an attractive (repulsive) force.
For a genuinely magnetodielectric plate the situation is more complex.
As the coefficient $C_4$ monotoneously decreases as a function of
$\varepsilon(0)$ and monotoneously increases as a function of
$\mu(0)$,
\begin{equation}
\label{eq91}
\frac{\partial C_4}{\partial\varepsilon(0)}<0, \quad
 \frac{\partial C_4}{\partial\mu(0)}>0, \quad
\end{equation}
the border between the attractive and repulsive potential, i.e., $C_4$
$\!=$ $\!0$, can be marked by a unique curve in the
$\varepsilon(0)\mu(0)$-plane, which is displayed in Fig.~\ref{fig5}.
In the limits of weak and strong magnetodielectric properties the
integral in Eq.~(\ref{eq87}) can be evaluated analytically. In the
case of weak magnetodielectric properties, 
\mbox{$\chi_\mathrm{e}(0)$ $\!\equiv$ $\!\varepsilon(0)$ $\!-1$
$\!\ll$ $\!1$} and $\chi_\mathrm{m}(0)$ $\!\equiv$ $\!\mu(0)$ $\!-$
$\!1$ $\!\ll$ $\!1$, the linear expansions
\begin{align}
\label{eq92}
&\frac{\varepsilon(0)v-\sqrt{\varepsilon(0)\mu(0)-1 + v^2 }}
 {\varepsilon(0)v+\sqrt{\varepsilon(0)\mu(0)- 1 + v^2 }}
 \nonumber\\
&\qquad\simeq
 \left[\frac{1}{2}-\frac{1}{4v^2}\right]\chi_\mathrm{e}(0)
 -\frac{1}{4v^2}\chi_\mathrm{m}(0)
\end{align}
and
\begin{align}
\label{eq93}
&\frac{\mu(0)v-\sqrt{\varepsilon(0)\mu(0)-1 + v^2}}
 {\mu(0)v+\sqrt{\varepsilon(0)\mu(0)-1 + v^2 }}
 \nonumber\\
&\qquad\simeq
 -\frac{1}{4v^2}\chi_\mathrm{e}(0)
 +\left[\frac{1}{2}-\frac{1}{4v^2}\right]\chi_\mathrm{m}(0)
\end{align}
lead to
\begin{equation}
\label{eq94}
C_4=-\frac{\hbar c\alpha_0^{(0)}(0)}{640\pi^2\varepsilon_0}
\bigl[23\ \chi_\mathrm{e}(0)-7\chi_\mathrm{m}(0)\bigr].
\end{equation}
For strong magnetodielectric properties, i.e., $\varepsilon(0)$
$\!\gg$ $\!1$ and $\mu(0)$ $\!\gg$ $\!1$, we may approximately set, on
noting that large values of $v$ are effectively suppressed in the
integral in Eq.~(\ref{eq87}),
\begin{equation}
\label{eq95}
\sqrt{\varepsilon(0)\mu(0)-1 + v^2 }\simeq
 \sqrt{\varepsilon(0)\mu(0)}\,,
\end{equation}
thus
\begin{align}
\label{eq96}
C_4 =& -\frac{3\hbar c\alpha_0^{(0)}(0)}{64\pi^2\varepsilon_0}
 \biggl[-\,\frac{2}{Z^3}\mathrm{ln}(1\!+\!Z)
 +\frac{2}{Z^2}+\frac{4}{Z}\mathrm{ln}(1\!+\!Z)\nonumber\\
& - \frac{1}{Z}-\frac{4}{3}-Z+2Z^2-2Z^3
 \mathrm{ln}\biggl(1+\frac{1}{Z}\biggr)\biggr],
\end{align}
with $Z$ $\!\equiv$ $\!\sqrt{\mu(0)/\varepsilon(0)}$ denoting the
static impedance of the material. Setting $C_4$ $\!=$ $\!0$ in
Eqs.~(\ref{eq94}) and (\ref{eq96}), we obtain the asymptotic
behaviour of the border curve in the two limiting cases. The result
shows that a repulsive vdW potential can be realized if
$\chi_\mathrm{m}(0)/\chi_\mathrm{e}(0)$ $\!\ge$ $\!23/7$ 
$\!=$~$\!3.29$ in the case of weak magnetodielectric properties, and
\mbox{$\mu(0)/\varepsilon(0)$ $\!\ge$ $\!5.11$} ($Z$ $\!\ge$
$\!2.26$) in the case of strong magnetodielectric properties. 

Apart from the different distance laws, the short-dis\-tan\-ce vdW
potential, Eq.~(\ref{eq88}), differs from the long-dis\-tan\-ce
potential, Eq.~(\ref{eq86}), in two respects. First, the relevant
coefficients $C_3$ and $C_1$ are not only determined by the static
values of the permittivity and the permeability, as is seen from
Eqs.~(\ref{eq89}) and (\ref{eq90}), and second,
Eqs.~(\ref{eq88})--(\ref{eq90}) reveal that electric and magnetic
properties give rise to potentials with different distance laws and
signs [\mbox{$C_3$ $\!>$ $\!0$} dominant (and $C_1$ $\!>$ $\!0$) if
\mbox{$\varepsilon$ $\!\neq$ $\!1$} and $\mu$ $\!=$ $\!1$, while
\mbox{$C_3$ $\!=$ $\!0$} and $C_1$ $\!>$ $\!0$ if \mbox{$\varepsilon$
$\!=$ $\!1$} and $\mu$ $\!\neq$ $\!1$]. However, although for the case
of a purely magnetic plate a repulsive vdW potential proportional to
$1/z_\mathrm{A}$ is predicted, in practice the attractive
$1/z_\mathrm{A}^3$ term will always dominate for sufficiently small
values of $z_\mathrm{A}$, because of the always existing electric
properties of the plate. Hence when in the long-distance limit the
potential becomes repulsive due to sufficiently strong magnetic
properties, then the formation of a potential wall at intermediate
distances becomes possible. It is evident that with decreasing
strength of the electric properties the maximum of the wall is shifted
to smaller distances while increasing in height. 

In the limiting case of weak electric properties, i.e.,
$\omega_\mathrm{Pe}/\omega_\mathrm{Te}$ $\!\ll$ $\!1$
and  $\omega_\mathrm{Pe}/\omega_\mathrm{Pm}$ $\!\ll$ $\!1$
[recall Eqs.~(\ref{eq84}) and (\ref{eq85})] one can thus expect that
the wall is situated within the short-distance range, so that
Eqs.~(\ref{eq88})--(\ref{eq90}) can be used to determine both its
position and height. {F}rom Eq.~(\ref{eq88}) we find that the wall
maximum is at 
\begin{equation}
\label{eq97}
z_\mathrm{A}^\mathrm{max}=\sqrt{\frac{3C_3}{C_1}}
\end{equation}
and has a height of
\begin{equation}
\label{eq98}
U(z_\mathrm{A}^\mathrm{max}) 
 =\frac{2}{3}\sqrt{\frac{C_1^3}{3C_3}}\, .
\end{equation}
In order to evaluate the integrals in Eqs.~(\ref{eq89}) and
(\ref{eq90}) for the coefficients $C_3$ and $C_1$, respectively, let
us restrict our attention to the case of a two-level atom and
disregard absorption ($\gamma_\mathrm{e}$ $\!\simeq$ $\!0$, 
$\!\gamma_\mathrm{m}$ $\!\simeq$ $\!0$). Straightforward calculation
yields ($\omega_\mathrm{Pe}/\omega_\mathrm{Te}$ $\!\ll$
$\!1$, $\omega_\mathrm{Pe}/\omega_\mathrm{Pm}$ $\!\ll$ $\!1$)
\begin{equation}
\label{eq99}
C_3\simeq\frac{|\vec{d}_{01}|^2}{96\pi\varepsilon_0}\,
 \frac{\omega_\mathrm{Pe}^2}{\omega_\mathrm{Te}^2}\,
 \frac{\omega_\mathrm{Te}}{\omega_{10}\!+\!\omega_\mathrm{Te}}
\end{equation} 
and
\begin{align}
\label{eq100}
C_1 \simeq&  \,\frac{\mu_0\hbar}{16\pi^2}
 \int_0^\infty\mathrm{d}u\ u^2\alpha_0^{(0)}(iu)
 \left[ \frac{\mu(iu)\!-\!1}{\mu(iu)\!+\!1} 
+ \frac{\mu(iu)\!-\!1}{2}\right]\nonumber\\
= & \,\frac{\mu_0|\vec{d}_{01}|^2\omega_\mathrm{Pm}^2}{96\pi}\,
 \frac{\omega_\mathrm{10}
 (2\omega_\mathrm{10}+\omega_\mathrm{Sm}+\omega_\mathrm{Tm})}
 {(\omega_\mathrm{10}+\omega_\mathrm{Sm})
 (\omega_\mathrm{10}+\omega_\mathrm{Tm})}
\end{align}
[$\omega_\mathrm{Sm}$ $\!=\!$ $\!(\omega_\mathrm{Tm}^2$ $\!+$
$\!\frac{1}{2}\omega_\mathrm{Pm}^2)^{1/2}$]. Substitution of
Eqs.~(\ref{eq99}) and (\ref{eq100}) into Eqs.~(\ref{eq97}) and
(\ref{eq98}), respectively, leads to 
\begin{align}
\label{eq101}
z_\mathrm{A}^\mathrm{max} =&\, 
 \frac{c}{\omega_\mathrm{Pm}}
 \frac{\omega_\mathrm{Pe}}{\omega_\mathrm{Te}}
 \sqrt{\frac{\omega_\mathrm{Te}(\omega_{10}+\omega_\mathrm{Tm})}
 {\omega_{10}(\omega_{10}+\omega_\mathrm{Te})}}
 \nonumber\\
&\quad\times\,\sqrt{\frac
 {3(\omega_{10}\!+\!\omega_\mathrm{Sm})}
 {(2\omega_\mathrm{10}+\omega_\mathrm{Sm}
 +\omega_\mathrm{Tm})}}
\end{align}
and
\begin{align}
\label{eq102}
U(z_\mathrm{A}^\mathrm{max}) 
=& \,\frac{|\mathbf{d}_{01}|^2
 \omega_\mathrm{Pm}^3}
 {48\pi \varepsilon_0 c^3}\,
 \frac{\omega_\mathrm{Te}}{\omega_\mathrm{Pe}}
 \sqrt{\frac{\omega_{10}+\omega_\mathrm{Te}}
 {\omega_\mathrm{Te}}}
 \nonumber\\
&\quad\times\,\left[\frac{\omega_{10}(2\omega_{10}
 \!+\!\omega_\mathrm{Sm}+\omega_\mathrm{Tm})}
 {3(\omega_{10}+\omega_\mathrm{Sm})
 (\omega_{10}+\omega_\mathrm{Tm})}\right]^{\frac{3}{2}}\, .
\end{align}
Note that consistency with the assumption of the wall occurring at
short distances requires that $z_\mathrm{A}^\mathrm{max}$ $\!\ll$
$c/\omega_\mathrm{M}^+$---a condition which is easily fulfilled for
sufficiently small values of $\omega_\mathrm{Pe}/\omega_\mathrm{Pm}$.
Inspection of Eq.~(\ref{eq102}) shows that the height of the wall
increases with $\omega_\mathrm{Pm}$, but decreases with increasing
$\omega_\mathrm{Tm}$ or increasing
$\omega_\mathrm{Pe}/\omega_\mathrm{Te}$ $\!=$
$\!\sqrt{\varepsilon(0)-1}$. Since the dependence of
$U(z_\mathrm{A}^\mathrm{max})$ on $\omega_\mathrm{Pm}$ is seen 
to be much stronger than its dependence on $\omega_\mathrm{Tm}$,
the wall height increases with $\omega_\mathrm{Tm}$ for given
$\omega_\mathrm{Pm}/\omega_\mathrm{Tm}$ $\!=$ $\!\sqrt{\mu(0)-1}$.

\begin{figure}[!t!]
\noindent
\begin{center}
\includegraphics[width=\linewidth]{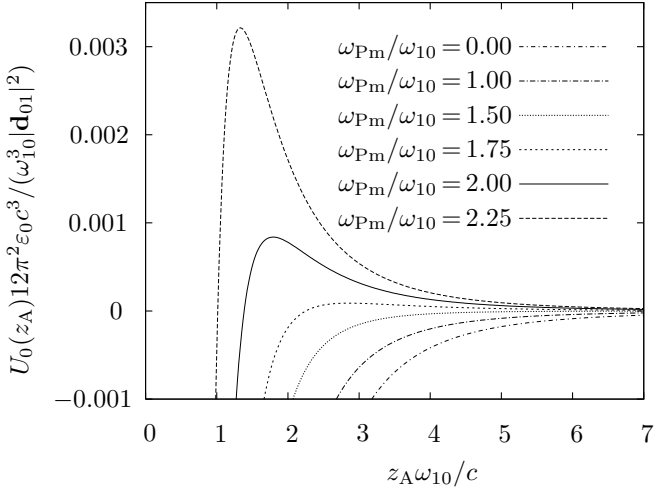}
\end{center}
\caption{
\label{fig2}
The vdW potential of a ground-state two-level atom situated in front
of an infinitely thick magnetodielectric plate is shown as a function
of the distance between the atom and the plate for different values of
$\omega_\mathrm{Pm}$ ($\omega_\mathrm{Pe}/\omega_{10}$ $\!=$ $\!0.75$,
$\omega_\mathrm{Te}/\omega_{10}$ $\!=$ $\!1.03$,
$\omega_\mathrm{Tm}/\omega_{10}$ $\!=$ $\!1$,
$\gamma_\mathrm{e}/\omega_{10}$ 
$\!=$ $\!\gamma_\mathrm{m}/\omega_{10}$ $\!=$ $\!0.001$).
}
\end{figure}

The distance-dependence of the vdW potential, as calculated
from Eq.~(\ref{eq81}) together with Eqs.~(\ref{eq82}) and (\ref{eq83})
for a two-level atom in front of a thick magnetodielectric plate whose
permittivity and permeability are modelled by Eqs.~(\ref{eq84}) and
(\ref{eq85}), respectively, is illustrated in Fig.~\ref{fig2}. The
figure reveals that the results derived above for the case where the
potential wall is observed in the short-distance range remain
qualitatively valid also for larger distances. So it is seen that for
sufficiently large values of $\omega_\mathrm{Pm}$ a potential wall
begins to form and grows in height as $\omega_\mathrm{Pm}$ increases.

In view of left-handed materials (cf.
Refs.~\cite{Pendry99,Smith00,Veselago68}), which simultaneously
exhibit negative real parts of $\varepsilon(\omega)$ and $\mu(\omega)$
within some (real) frequency interval such that the real part of the
refractive index becomes negative therein, the question may arise
whether these materials would have an exceptional effect on the
ground-state CP force. The answer is obviously no, because the
ground-state vdW potential as given by Eq.~(\ref{eq81}) together with
Eqs.~(\ref{eq82}) and (\ref{eq83}) is expressed in terms of the always
positive values of the permittivity and the permeability at imaginary
frequencies. Clearly, the situation may change for an atom prepared in
an excited state. In such a case, the vdW potential is essentially
determined by the real part of the Green tensor [cf. Eqs.~(\ref{eq63})
and (\ref{eq65})]. When there are transition frequencies that lie in
frequency intervals where the material behaves left-handed, then
particularities may occur.


\subsection{Plate of finite thickness}
\label{sec4.2}

\begin{figure}[!t!]
\noindent
\begin{center}
\includegraphics[width=\linewidth]{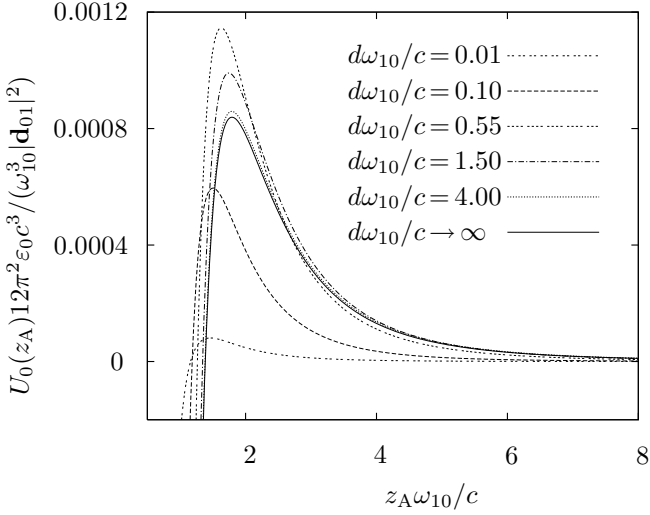}
\end{center}
\caption{
\label{fig6}
The vdW potential of a ground-state two-level atom situated in front
of a magnetodielectric plate is shown as a function of the distance
between the body and the interface for different values of the plate
thickness $d$ 
($\omega_\mathrm{Pe}/\omega_{10}$ $\!=$ $\!0.75$,
$\omega_\mathrm{Te}/\omega_{10}$ $\!=$ $\!1.03$,
$\omega_\mathrm{Pm}/\omega_{10}$ $\!=$ $\!2$,
$\omega_\mathrm{Tm}/\omega_{10}$ $\!=$ $\!1$,
$\gamma_\mathrm{e}/\omega_{10}$ 
$\!=$ $\!\gamma_\mathrm{m}/\omega_{10}$ $\!=$~$\!0.001$).
}
\end{figure}

Let us now consider an atom in front of a magnetodielectric plate
of finite thickness $d_1$ $\!\equiv\!$ $d$
[$n$ $\!=$ $\!j$ $\!=$ $\!2$, $\varepsilon_1(\omega)$ $\!\equiv$ 
$\!\varepsilon(\omega)$, $\mu_1(\omega)$ $\!\equiv$ $\!\mu(\omega)$,
$\varepsilon_0(\omega)$ $\!=$ $\!\varepsilon_2(\omega)$ $\!\equiv$ 
$\!1$, $\mu_0(\omega)$ $\!=$ $\!\mu_2(\omega)$ $\!\equiv$ $\!1$].
Using Eqs.~(\ref{eq73}) and (\ref{eq74}) we find that the reflection
coefficients in Eq.~(\ref{eq81}) are now given by ($b_1$ $\!\equiv$
$\!b_\mathrm{M}$)
\begin{eqnarray}
\label{eq103}
r_{n-}^s
&\!=&\!\frac{[\mu^2(iu)b^2-b_\mathrm{M}^2]\tanh(b_\mathrm{M}d)}
 {2\mu(iu)b b_\mathrm{M}+[\mu^2(iu)b^2+b_\mathrm{M}^2]
 \tanh(b_\mathrm{M}d)}\,,\quad\\
\label{eq104}
r_{n-}^p
&\!=&\!
 \frac{[\varepsilon^2(iu)b^2-b_\mathrm{M}^2]\tanh(b_\mathrm{M}d)}
 {2\varepsilon(iu)b b_\mathrm{M}+
 [\varepsilon^2(iu)b^2+b_\mathrm{M}^2]\tanh(b_\mathrm{M}d)}\,.
\end{eqnarray}
Typical examples of the vdW potential obtained by numerical evaluation
of Eq.~(\ref{eq81}) [together with Eqs.~(\ref{eq103}) and
(\ref{eq104})] for a two-level atom are shown in Fig.~\ref{fig6},
revealing that for sufficiently strong magnetic properties the
formation of a repulsive potential wall can also be observed for a  
magnetodielectric plate of finite thickness. In the figure, the medium
parameters correspond to those which have already been found to
support the formation of a repulsive potential wall in the case of an
infinitely thick plate. We see that the qualitative behaviour of the
vdW potential is independent of the layer thickness. In particular,
all curves in Fig.~\ref{fig6} feature a repulsive long-range potential
that leads to a potential wall of finite height, the potential
becoming attractive at very short distances. However, the position and
height of the wall are seen to vary with the thickness of the plate.
While the position of the wall shifts only slightly as the plate
thickness is changed from very small to very large values, the height
of the wall reacts very sensitively as the plate thickness is varied.
For small values of the thickness the potential height is very small,
it increases towards a maximum, and then decreases asymptotically
towards the value found for the infinitely thick plate as the
thickness is increased further towards very large values. It is worth
noting that there is an optimal plate thickness for creating a maximum
potential wall. In this case the plate thickness is comparable to the
position of the potential maximum---a case which is realized between
the two extremes of infinitely thick and infinitely thin layer
thickness. 

Further insight can be gained by considering the two limiting cases of
an infinitely thick and an asymptotically thin plate. It is obvious
that the integration in Eq.~(\ref{eq81}) is effectively limited by the
exponential factor $e^{-2bz_\mathrm{A}}$ to a circular region where
\mbox{$b$ $\!\lesssim$ $\!1/(2z_\mathrm{A})$}. In particular, in the
limit of a sufficiently thick plate, \mbox{$d$ $\!\gg$
$\!z_\mathrm{A}$}, the estimate
\begin{equation}
\label{eq105}
b_\mathrm{M}d\ge bd\sim\frac{d}{2z_\mathrm{A}}\gg 1
\end{equation}
[recall Eqs.~(\ref{eq75}) and (\ref{eq77})] is valid within (the major
part of) the effective region of integration, and one may hence make
the approximation $\tanh(b_\mathrm{M}d)$ $\!\simeq$ $\!1$ in
Eqs.~(\ref{eq103}) and (\ref{eq104}), which then obviously reduce to
Eqs.~(\ref{eq82})  and (\ref{eq83}) valid for an infinitely thick
plate. On the contrary, in the limit of an asymptotically thin plate,
$\sqrt{\varepsilon(0)\mu(0)}d$ $\!\ll$ $\!z_\mathrm{A}$, we find that
the inequalities
\begin{align}
\label{eq106}
b_\mathrm{M}d
&\le \sqrt{\varepsilon(iu)\mu(iu)}\,bd
\le \sqrt{\varepsilon(0)\mu(0)}\,bd
\nonumber\\
&\le\frac{\sqrt{\varepsilon(0)\mu(0)}d}{2z_\mathrm{A}}
\ll 1
\end{align}
hold in the effective region of integration, and one may hence
linearly expand the integrand in Eq.~(\ref{eq81}) in terms of
$b_\mathrm{M}d$, which is equivalent to approximating the reflection
coefficients (\ref{eq103}) and (\ref{eq104}) according to 
\begin{eqnarray}
\label{eq107}
r_{n-}^s
&\simeq&\frac{\mu^2(iu)b^2-b_\mathrm{M}^2} 
 {2\mu(iu)b}\,d\,,\\
\label{eq108}
r_{n-}^p
&\simeq&\frac{\varepsilon^2(iu)b^2-b_\mathrm{M}^2}
 {2\varepsilon(iu)b}\,d\,.
\end{eqnarray}
As in the case of an infinitely thick plate, cf. Sec.~\ref{sec4.1},
the dependence of the vdW potential on the atom-plate separation in
the case of an asymptotically thin plate reduces to simple power laws 
in the long- and short-distance limits. In the long-distance limit,
\mbox{$z_\mathrm{A}$ $\!\gg$
$\!c/\omega_\mathrm{A}^-$, $z_\mathrm{A}$
$\!\gg$ $\!c/\omega_\mathrm{M}^-$},
Eq.~(\ref{eq81}) together with Eqs.~(\ref{eq107}) and (\ref{eq108})
reduces to (see Appendix \ref{appB}) 
\begin{equation}
\label{eq109}
U(z_\mathrm{A})=\frac{D_5}{z_\mathrm{A}^5}\, ,
\end{equation} 
where
\begin{equation}
\label{eq110}
D_5=-\frac{\hbar c\alpha_0^{(0)}(0)d}{160\pi^2\varepsilon_0}\,
 \biggl[\frac{14\varepsilon^2(0)-9}{\varepsilon(0)}
 -\frac{6\mu^2(0)-1}{\mu(0)}\biggr]\, ,
\end{equation}
while in the short-distance limit,
$z_\mathrm{A}$ $\!\ll$ $\!c/\omega_\mathrm{A}^+$ and/or
$z_\mathrm{A}$ $\!\ll$~$\!c/\omega_\mathrm{M}^+$, 
Eq.~(\ref{eq81}) together with Eqs.~(\ref{eq107}) and (\ref{eq108})
can be approximated by (see Appendix \ref{appB}) 
\begin{equation}
\label{eq111}
U(z_\mathrm{A})=-\frac{D_4}{z_\mathrm{A}^4}
 +\frac{D_2}{z_\mathrm{A}^2}\, ,
\end{equation}
where
\begin{equation}
\label{eq112}
D_4= \frac{3\hbar d}{64\pi^2\varepsilon_0}
 \int_0^\infty\mathrm{d}u\,
 \alpha_0^{(0)}(iu)\frac{\varepsilon^2(iu)-1}{\varepsilon(iu)}\ge 0
\end{equation}
and
\begin{align}
\label{eq113}
&D_2 = \frac{\mu_0\hbar d}{64\pi^2}
 \int_0^\infty\mathrm{d}u\,u^2
 \alpha_0^{(0)}(iu)\Biggl\{\frac{\varepsilon^2(iu)-1}{\varepsilon(iu)}
 \nonumber\\&\hspace{6ex}
+\frac{\mu^2(iu)-1}{\mu(iu)}
 +\frac{2[\varepsilon(iu)\mu(iu)-1]}{\varepsilon(iu)}\Biggr\}\ge 0\,.
\end{align}

Comparing the power laws (\ref{eq109}) and (\ref{eq111}) with
those obtained for an infinitely thick plate, Eqs.~(\ref{eq86}) and
(\ref{eq88}), we see that the powers of $1/z_\mathrm{A}$ are
universally increased by one. Again, we find that in the long-distance
limit the vdW potential follows a power law that is independent of the
material properties of the plate, the sign being determined by the
relative strengths of the magnetic and electric properties (a purely
electric plate creates an attractive vdW potential, while a purely
magnetic plate gives rise to a repulsive one). And again the
short-distance behaviours of the vdW potential for plates of different
material properties (i.e., electric/magnetic) differ in both sign and
leading power law (the repulsive potential in the case of a purely
magnetic plate being weaker than the attractive potential in the case
of a purely electric plate by two powers in the atom-plate
separation). Interestingly, a similar behaviour, i.e., the same
hierarchy of power laws and the same signs have been found for the vdW
force between two atoms \cite{Sucher68,Boyer69,Farina02}
and for the Casimir force between two semi-infinite half spaces
\cite{Henkel04}. This is illustrated in Tab.~\ref{tab1}, where the
asymptotic power laws found for an atom interacting with an infinitely
thick plate, Eqs.~(\ref{eq86}) and (\ref{eq90}), and an asymptotically
thin plate, Eqs.~(\ref{eq109}) and (\ref{eq113}), are summarized and
compared to those valid for the interactions between two atoms or two
half spaces, respectively.  
\begin{table}
\begin{center}
 \begin{tabular}{|c||c|c|c|c|}
\hline
 distance&\multicolumn{2}{c|}{short}&\multicolumn{2}{c|}{long}\\ 
\hline
 polarizability&$\mathrm{e}\leftrightarrow \mathrm{e}$
 &$\mathrm{e}\leftrightarrow \mathrm{m}$
 &$\mathrm{e}\leftrightarrow \mathrm{e}$
 &$\mathrm{e}\leftrightarrow \mathrm{m}$\\ 
\hline\hline
 atom $\leftrightarrow$ h.s.
 &\parbox{5ex}{$$-\frac{1}{z^4}$$}
 &\parbox{5ex}{$$+\frac{1}{z^2}$$}
 &\parbox{5ex}{$$-\frac{1}{z^5}$$}
 &\parbox{5ex}{$$+\frac{1}{z^5}$$}\\
\hline
 atom $\leftrightarrow$ thin plate
 &\parbox{5ex}{$$-\frac{1}{z^5}$$}
 &\parbox{5ex}{$$+\frac{1}{z^3}$$}  
 &\parbox{5ex}{$$-\frac{1}{z^6}$$}
 &\parbox{5ex}{$$+\frac{1}{z^6}$$}\\ 
\hline
 atom $\leftrightarrow$ atom 
 &\parbox{5ex}{$$-\frac{1}{z^7}$$}
 &\parbox{5ex}{$$+\frac{1}{z^5}$$}  
 &\parbox{5ex}{$$-\frac{1}{z^8}$$}
 &\parbox{5ex}{$$+\frac{1}{z^8}$$}\\ 
\hline
 h.s. $\leftrightarrow$ h.s. 
 &\parbox{5ex}{$$-\frac{1}{z^3}$$}
 &\parbox{5ex}{$$+\frac{1}{z}$$}  
 &\parbox{5ex}{$$-\frac{1}{z^4}$$}
 &\parbox{5ex}{$$+\frac{1}{z^4}$$}\\ 
\hline
\end{tabular}
\end{center}
\caption{
\label{tab1}
Signs and asymptotic power laws of the forces between various
polarizable objects. In the table heading, $\mathrm{e}$ stands for a
purely electric object and $\mathrm{m}$ for a purely magnetic one. The
signs $+$ and $-$ denote repulsive and attractive forces,
respectively. Half space is abbreviated by h.s..}
\end{table}

For weak magnetodielectric properties, the similarity of the results
displayed in Tab.~\ref{tab1} can be regarded as being a consequence of
the additivity of vdW-type interactions. In fact, in this case (which
for a gaseous medium of given atomic species corresponds to a
sufficiently dilute gas) all results of the table can be derived from
the vdW interaction of two single atoms via pairwise summation. The
additivity can explicitly be seen when comparing the result found for
an asymptotically thin plate with that of an infinitely thick plate in
the case of weak magnetodielectric properties
[$\chi_\mathrm{e}(iu)$ $\!\equiv$ $\!\varepsilon(iu)$ $\!-1$
$\!\ll$ $\!1$, $\chi_\mathrm{m}(iu)$ $\!\equiv$ $\!\mu(iu)$ $\!-$
$\!1$ $\!\ll$~$\!1$]. Making a linear expansion in
$\chi_\mathrm{e}(iu)$ and $\chi_\mathrm{m}(iu)$, we find that the vdW
potential of an infinitely thick plate, Eq.~(\ref{eq81}) together with
Eqs.~(\ref{eq82}) and (\ref{eq83}), reduces to 
\begin{align}
\Delta_1U(z_\mathrm{A})
&= -\frac{\hbar\mu_0}{8\pi^2}
 \int_0^{\infty}\!\!\mathrm{d}u\,u^2 \alpha^{(0)}(iu)
 \int_0^\infty\!\!\mathrm{d}q\,\frac{q}{b}e^{-2bz_\mathrm{A}}
 \nonumber
\end{align} 
\begin{align}
\label{eq114}
&\hspace{12ex}\times\,\Biggl\{\Biggl[\biggl(\frac{bc}{u}
\biggr)^2-1+\frac{1}{2}
 \biggl(\frac{u}{bc}\biggr)^2\Biggr]\chi_\mathrm{e}(iu)
 \nonumber\\
&\hspace{18ex}-\Biggl[1-\frac{1}{2}\biggl(\frac{u}{bc}\biggr)^2\Biggr]
 \chi_\mathrm{m}(iu)\Biggr\}\,,
\end{align}
while the vdW potential of an asymptotically thin plate,
Eq.~(\ref{eq81}) together with Eqs.~(\ref{eq107}) and (\ref{eq108}),
can be approximated by
\begin{align}
\label{eq115}
\Delta_1U^d(z_\mathrm{A})
&=-\frac{\hbar\mu_0d}{4\pi^2}
 \int_0^{\infty}\!\!\mathrm{d}u\,u^2 \alpha^{(0)}(iu)
 \int_0^\infty\!\!\mathrm{d}q\,qe^{-2bz_\mathrm{A}}
 \nonumber\\
&\hspace{2ex}\times\,\Biggl\{\Biggl[\biggl(\frac{bc}{u}\biggr)^2-1
+\frac{1}{2}\biggl(\frac{u}{bc}\biggr)^2\Biggr]\chi_\mathrm{e}(iu)
 \nonumber\\
&\hspace{8ex}-\Biggl[1-\frac{1}{2}\biggl(\frac{u}{bc}\biggr)^2\Biggr]
 \chi_\mathrm{m}(iu)\Biggr\}\,.
\end{align}
Comparison of Eqs.~(\ref{eq114}) and (\ref{eq115}) shows that for
weakly magnetodielectric media the vdW potential of an infinitely
thick plate is simply the integral over an infinite number of
thin-plate vdW potentials,
\begin{equation}
\label{eq116}
U_0(z_\mathrm{A})=
\int_{z_\mathrm{A}}^\infty\frac{\mathrm{d}z}{d}\,U^d_0(z).
\end{equation}
In the case of media with stronger magnetodielectric properties
many-body interactions may be thought of as preventing the vdW
potential from being additive so that a relation of the type of
Eq.~(\ref{eq116}) is not true in general. As a consequence, the
coefficients of the asymptotic power laws in Tab.~\ref{tab1} can not
be related to each other via simple additivity arguments in general.
However, we note from Tab.~\ref{tab1} that the consideration of
many-body corrections only changes the coefficients of the asymptotic
power laws, not the power laws themselves.


\subsection{Planar cavity}
\label{sec4.3}

Finally, let us consider an atom placed within the simplest type of
planar cavity, i.e., between two identical infinitely thick
magnetodielectric plates which are separated by a distance $d_1$
$\!\equiv$ $\!s$ [$n$ $\!=$ $\!2$, $j$ $\!=$ $\!1$, 
$\varepsilon_1(\omega)$ $\!=$ $\!\mu_1(\omega)$ $\!\equiv$ $\!1$,
$\varepsilon_0(\omega)$ $\!=$ $\!\varepsilon_2(\omega)$ $\!\equiv$
$\varepsilon(\omega)$, $\mu_0(\omega)$ $\!=$ $\!\mu_2(\omega)$
$\!\equiv$ $\!\mu(\omega)$]. 
{F}rom Eqs.~(\ref{eq73}) and (\ref{eq74}) it then follows that the
reflection coefficients in Eq.~(\ref{eq80}) are given by
($b_0$ $\!=$ $\!b_2$ $\!\equiv$ $b_\mathrm{M}$)
\begin{align}
\label{eq117}
&r_{j-}^s = r_{j+}^s=
 \frac{\mu(iu)b-b_\mathrm{M}}{\mu(iu)b+b_\mathrm{M}}\,,\\
\label{eq118} 
&r_{j-}^p = r_{j+}^p=
 \frac{\varepsilon(iu)b-b_\mathrm{M}}{\varepsilon(iu)b+b_\mathrm{M}}
 \,.
\end{align}

\begin{figure}[!t!]
\noindent
\begin{center}
\includegraphics[width=\linewidth]{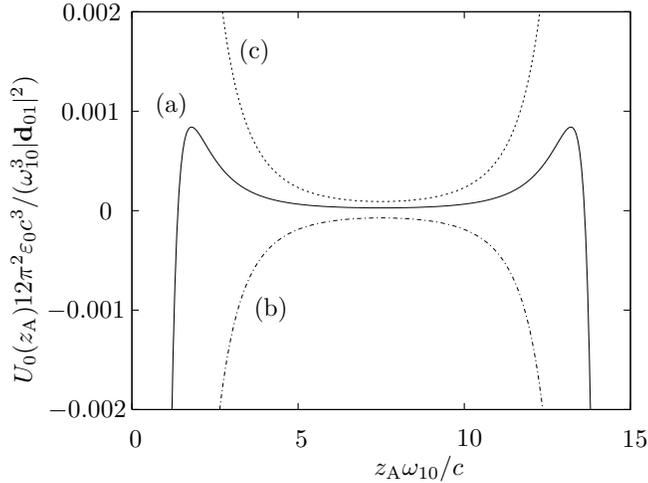}
\end{center}
\caption{
\label{fig7}
The vdW potential of a ground-state two-level atom situated between 
two infinitely thick (a) magnetodielectric plates
($\omega_\mathrm{Pe}/\omega_{10}$ $\!=$ $\!0.75$,
$\omega_\mathrm{Te}/\omega_{10}$ $\!=$ $\!1.03$,
\mbox{$\omega_\mathrm{Pm}/\omega_{10}$ $\!=$ $\!2$},
\mbox{$\omega_\mathrm{Tm}/\omega_{10}$ $\!=$ $\!1$},
$\gamma_\mathrm{e}/\omega_{10}$ 
$\!=$ $\!\gamma_\mathrm{m}/\omega_{10}$ $\!=$ $\!0.001$)
(b) dielectric plates [$\mu(\omega)$ $\!\equiv$ $\!1$, other
parameters as in (a)], (c) magnetic plates [$\varepsilon(\omega)$
$\!\equiv$ $\!1$, other parameters as in (a)], which are separated by
a distance \mbox{$s$ $\!=$ $\!15c/\omega_{10}$}, is shown as a
function of the position of the atom.
}
\end{figure}
Examples of the vdW potential of a two-level atom between two
identical infinitely thick magnetodieletric plates as calculated from
Eq.~(\ref{eq80}) together with Eqs.~(\ref{eq117}) and (\ref{eq118})
are plotted in Fig.~\ref{fig7}. It is seen that the attractive
(repulsive) potentials associated with each of two purely electric
(magnetic) plates combine to an infinite potential wall (well) at the
center of the cavity. Hence, a potential well of finite depth at the
center of the cavity can be realized in the case of two genuinely
magnetodielectric plates of sufficiently strong magnetic properties as
shown in the figure. Provided that appropriate materials would be
available, this feature could in principle be used for the trapping 
and guiding of atoms.


\section{Summary and Conclusions}
\label{sec5}

Within the framework of exact macroscopic QED in linear, causal media,
we have given a unified theory of the CP force acting on an atom when
placed near an arbitrary arrangement of dispersing and absorbing
magnetodielectric bodies. We have considered both the familiar
perturbative approach to the problem, where the atom-field coupling
energy calculated in lowest-order perturbation theory is regarded as
the potential associated with the CP force acting on the atom prepared
in an energy eigenstate, and a dynamical approach based on the Lorentz
force averaged with respect to the body-assisted electromagnetic
vacuum and the internal motion of the atom. In particular, the theory
allows to extend the quantum mechanical calculation of the interaction
energy to the realistic case of material dispersion and absorption---a
case for which standard mode expansion of the electromagnetic field
runs into difficulties. So, the theory yields the vdW potential in
terms of the electromagnetic-field scattering Green tensor and the
lowest-order atomic polarizability in a natural manner, without
borrowing arguments from other theories such as the widely used linear
response theory.  

In contrast to the perturbative treatment of the CP force, the
dynamical treatment allows for including arbitrary excited atomic
states, their temporal evolution and thus transient components of the
force, and the influence of the body-induced shifting and broadening
of the atomic transitions on the force. Whereas level shifting can, at
least for very small atom-body distances, noticeably modify both the
resonant and the off-resonant force components, level broadening
effectively affects only the resonant components. Thus the
perturbative treatment may be justified for the purely off-resonant
ground-state force, while being inadequate for the excited-state force
containing resonant components (leaving aside its obvious inablity to
describe the transient nature of excited-state components).
  
Finally, we have applied the theory to analyze the competing effects
of the electric and magnetic properties on the CP force acting on a
ground-state atom placed within a magnetodielectric multilayer system,
studying the corresponding vdW potential for the cases of thick and
thin plates as well as a planar cavity. In close analogy to the vdW
interaction between two atoms or the Casimir force between two plates,
the electric and magnetic properties compete in creating attractive
and repulsive force components, respectively. In particular, if the
atom interacts with a magnetodielectric plate of sufficiently strong
magnetic properties, a potential wall can be formed. We have given
conditions for the creation of such a wall and shown that there is an
optimal plate thickness for maximizing the height of the wall. Placing
the atom between two magnetodielectric plates each of which giving
rise to a potential wall, one can combine the two potentials to a
potential well. Needless to say that the thorough understanding of the
interplay of electric and magnetic material properties can serve as a
roadmap showing desirable directions of research in material design
when aiming at shaping vdW potentials in a controlled way. 


\begin{acknowledgement}
We thank J. B. Pendry for valuable discussions. 
This work was supported by the Deutsche Forschungsgemeinschaft.
S.Y.B. is grateful for having been granted a Th\"{u}\-rin\-ger
Landesgraduier\-tenstipendium and acknowledges support by the E.W.
Kuhl\-mann-Foundation. T.K. is grateful for being member of
Gradu\-ierten\-kolleg 567, which is funded by the Deutsche
Forschungsgemeinschaft and the Government of
Mecklenburg-Vorpom\-mern. 
\end{acknowledgement}


\appendix

\section{Long- and short-distance limits}
\label{appB}

The long-distance (short-distance) limit corresponds to separation
distances $z_\mathrm{A}$ between the atom and the multilayer system
which are much greater (smaller) than the wavelenghts corresponding to
typical frequencies of the atom and the multilayer system. To obtain
approximate results for the two limiting cases, let us analyze the
$u$-integral in Eq.~(\ref{eq81}) in a little more detail and begin
with the long-distance limit, i.e.,
\begin{align}
\label{B1}
z_\mathrm{A}\gg 
 \frac{c}{\omega_\mathrm{A}^-}\,,\quad 
 z_\mathrm{A}\gg \frac{c}{\omega_\mathrm{M}^-}\,,
\end{align}
where $\omega_\mathrm{A}^-$ $\!=$ $\!\mathrm{min}(\{\omega_{k0}|k$
$\!=$ $\!1,2\ldots\})$ is the lowest atomic transition frequency, and 
$\omega_\mathrm{M}^-$ $\!=$ 
$\!\mathrm{min}(\omega_\mathrm{Te},\omega_\mathrm{Tm})$
is the lowest medium resonance frequency. For convenience, we
introduce the new integration variable 
\begin{equation}
\label{B2}
v = \frac{cb}{u}
\end{equation}
and transform the integral according to
\begin{align}
\label{B3}
\int_0^\infty\mathrm{d}u &
 \int_0^\infty\mathrm{d}q\,\frac{q}{b}\,e^{-2bz_\mathrm{A}}\ldots
 \nonumber\\
&\mapsto\int_1^\infty\mathrm{d}v
 \int_0^\infty\mathrm{d}u\,
 \frac{u}{c}\,e^{-2z_\mathrm{A}vu/c}\ldots\ ,
\end{align}
where $b_\mathrm{M}$ has to be replaced according to 
\begin{equation}
\label{B4}
b_\mathrm{M}\mapsto
 \frac{u}{c}\,\sqrt{\varepsilon(iu)\mu(iu)-1+v^2}. 
\end{equation}
Inspection of Eqs.~(\ref{eq81}) together with Eqs.~(\ref{eq82}) and
(\ref{eq83}), or Eqs.~(\ref{eq107}) and (\ref{eq108}), respectively,
as well as Eq.~(\ref{B3}) reveals that the frequency interval giving
the main contribution to the respective $u$-integral is determined by
a set of effective cutoff functions, namely
\begin{equation}
\label{B5}
f(u)=e^{-2z_\mathrm{A}u/c},
\end{equation}
\begin{equation}
\label{B6}
g_k(u)=\frac{1}{1+(u/\omega_{k0})^2}\,,
\end{equation}
which enter via the atomic polarizability, cf. Eq.~(\ref{eq47}), and 
\begin{align}
\label{B7}
&h_\mathrm{e}(u)
 = \frac{1}{1+(u/\omega_\mathrm{Te})^2}\,,
 \\
\label{B8}
&h_\mathrm{m}(u)
 = \frac{1}{1+(u/\omega_\mathrm{Tm})^2}\,,
\end{align}
which enter the reflection coefficients as given by Eqs.~(\ref{eq82})
and (\ref{eq83}), or Eqs.~(\ref{eq107}) and (\ref{eq108}),
respectively, via $\varepsilon(iu)$ and $\mu(iu)$, cf.
Eqs.~(\ref{eq84}) and (\ref{eq85}). The cutoff functions obviously
give their main contributions in regions, where 
\begin{align}
\label{B9}
&\hspace{10ex} u\lesssim \frac{c}{2z_\mathrm{A}} & \mathrm{for} 
& \quad f(u),\hspace{10ex}\\
\label{B10}
&\hspace{10ex} u\lesssim\omega_{k0} & \mathrm{for} 
& \quad g_k(u),\hspace{10ex}\\
\label{B11}
&\hspace{10ex} u\lesssim\omega_\mathrm{Te} & \mathrm{for}
& \quad h_\mathrm{e}(u),\hspace{10ex}\\
\label{B12}
&\hspace{10ex} u\lesssim\omega_\mathrm{Tm} & \mathrm{for} 
& \quad h_\mathrm{m}(u).\hspace{10ex}
\end{align}
Combining Eqs.~(\ref{B9})--(\ref{B12}) with Eq.~(\ref{B1}), we find
that the function $f(u)$ effectively limits the $u$-integration to a
region where
\begin{align}
\label{B13}
&\frac{u}{\omega_{k0}} \le \frac{u}{\omega_\mathrm{A}^-}
 \lesssim \frac{c}{2z_\mathrm{A}\omega_\mathrm{A}^-} \ll 1,
 \\[1ex]
\label{B14}
&\frac{u}{\omega_\mathrm{Te}} \le \frac{u}{\omega_\mathrm{M}^-}
 \lesssim \frac{c}{2z_\mathrm{A}\omega_\mathrm{M}^-}\ll 1,
 \\[1ex]
\label{B15}
&\frac{u}{\omega_\mathrm{Tm}} \le \frac{u}{\omega_\mathrm{M}^-}
 \lesssim \frac{c}{2z_\mathrm{A}\omega_\mathrm{M}^-} \ll 1.
\end{align}
Performing a leading-order expansion of the integrand in
Eq.~(\ref{eq81}) in terms of the small quantities 
$u/\omega_{k0}$, $u/\omega_\mathrm{Te}$, and $u/\omega_\mathrm{Tm}$,
we may set 
\begin{equation}
\label{B16}
\alpha_0^{(0)}(iu)\simeq\alpha_0^{(0)}(0),\  
 \varepsilon(iu)\simeq\varepsilon(0),\ 
 \mu(iu)\simeq\mu(0).
\end{equation}
Combining  Eqs.~(\ref{B2})--(\ref{B4}) and Eq.~(\ref{B16})  
with Eq.~(\ref{eq81}) together with Eqs.~(\ref{eq82}) and
(\ref{eq83}), or Eqs.~(\ref{eq107}) and (\ref{eq108}), respectively,
and evaluating the remaining $u$-integrals we arrive at
Eq.~(\ref{eq86}) [together with Eq.~(\ref{eq87})] and
Eq.~(\ref{eq109}) [together with Eq.~(\ref{eq110})]. 

The short-distance limit, on the contrary, is defined by
\begin{align}
\label{B17}
z_\mathrm{A}\ll 
 \frac{c}{\omega_\mathrm{A}^+}\quad\mathrm{and/or}\quad
 z_\mathrm{A}\ll \frac{c}{\omega_\mathrm{M}^+}\,,
\end{align}
where $\omega_\mathrm{A}^+$ $\!=$ $\!\mathrm{max}(\{\omega_{k0}|k$
$\!=$ $\!1,2,\ldots\})$ is the highest inneratomic transition
frequency and $\omega_\mathrm{M}^+$ $\!=\!$ 
$\mathrm{max}(\omega_\mathrm{Te},\omega_\mathrm{Tm})$
is the highest medium resonance frequency. Again, it is convenient to
change the integration variables in Eq.~(\ref{eq81}), but now we
transform according to
\begin{align}
\label{B18}
\int_0^\infty\mathrm{d}u &
 \int_0^\infty\mathrm{d}q\,\frac{q}{b}\,e^{-2bz_\mathrm{A}}\ldots
 \nonumber\\
&\mapsto\int_0^\infty\mathrm{d}u\,
 \int_{u/c}^\infty\mathrm{d}b\,
 e^{-2bz_\mathrm{A}}\ldots\ ,
\end{align}
where $b_\mathrm{M}$ has to be replaced according to 
\begin{equation}
\label{B19}
b_\mathrm{M}\mapsto
 \sqrt{\frac{u^2}{c^2}\big[\varepsilon(iu)\mu(iu)-1\big] 
 +b^2}\ .
\end{equation} 
Combining Eqs.~(\ref{B9})--(\ref{B12}) with Eq.~(\ref{B17}) reveals
that the functions $g_k(u)$, $h_\mathrm{e}(u)$, and $h_\mathrm{m}(u)$
limit the $u$-integra\-tion to a region where
\begin{equation}
\label{B20}
\frac{z_\mathrm{A}u}{c}\lesssim
 \frac{z_\mathrm{A}\omega_\mathrm{A}^+}{c}\ll 1
\end{equation}
and/or
\begin{equation}
\label{B21}
\frac{z_\mathrm{A}u}{c}\lesssim
 \frac{z_\mathrm{A}\omega_\mathrm{M}^+}{c}\ll 1.
\end{equation}
A valid approximation to the $u$-integral in Eq.~(\ref{eq81}) can
hence be obtained by performing a Taylor exansion in 
$z_\mathrm{A}u/c$. To that end, we apply the transformation
(\ref{B18}) to Eq.~(\ref{eq81}) together with Eqs.~(\ref{eq82})
and (\ref{eq83}), or Eqs.~(\ref{eq107}) and (\ref{eq108}),
respectively, retain only the leading-order terms in $u/(cb)$
(corresponding to the leading-order terms in $z_\mathrm{A}u/c$ in the
$u$-integral) and carry out the $b$-integral. After again discarding
higher-order terms in $z_\mathrm{A}u/c$, we arrive at Eq.~(\ref{eq88})
[together with Eqs.~(\ref{eq89}) and (\ref{eq90})] and
Eq.~(\ref{eq111}) [together with Eqs.~(\ref{eq112}) and
(\ref{eq113})], respectively.


\end{document}